\documentclass[pra,twocolumn,superscriptaddress,longbibliography,floatfix,nofootinbib]{revtex4-2}
\usepackage[utf8]{inputenc}
\usepackage[T1]{fontenc}
\usepackage{amsmath,amssymb,amsfonts}
\usepackage{graphicx}
\usepackage{braket}
\usepackage{booktabs}
\usepackage{multirow}
\usepackage{makecell}
\usepackage{xcolor}
\usepackage{url}
\usepackage{hyperref}
\usepackage[capitalise]{cleveref}
\usepackage{enumitem}
\newcommand{\CC}{\mathcal{C}}
\newcommand{\DD}{\mathcal{D}}
\newcommand{\EE}{\mathcal{E}}
\newcommand{\UU}{\mathcal{U}}
\newcommand{\tr}{\mathrm{Tr}}
\newcommand{\Frec}{F_\mathrm{rec}}
\newcommand{\Fest}{F_\mathrm{est}}
\newcommand{\Fnoisy}{F_\mathrm{noisy}}
\newcommand{\rhot}{\rho_\mathrm{target}}
\newcommand{\rhon}{\rho_\mathrm{noisy}}
\newcommand{\rhor}{\rho_\mathrm{rec}}
\newcommand{\rhoe}{\hat{\rho}_\mathrm{est}}

\begin{document}

\title{Blind catalytic quantum error correction: target-state estimation and fidelity bounds}

\author{Hikaru Wakaura}
\email{h.wakaura@qiri.co.jp}
\affiliation{QIRI (Quantum Integrated Research Institute Inc.), 1-16-3, Akasaka, Minato-ku, Tokyo 107-0061, Japan}

\begin{abstract}
Near-term quantum computers must protect fragile coherence against decoherence to deliver useful results.
Catalytic quantum error correction (CQEC) addresses this by amplifying residual coherence with a reusable catalyst, achieving threshold-free recovery whenever the target coherent modes survive in the noisy state.
However, the original protocol requires complete knowledge of the ideal target---an assumption that fails for variational and iterative algorithms whose output is unknown to the correction module.
Here we show that this requirement can be removed by estimating the target from the noisy output alone, in a two-stage protocol we call \emph{blind CQEC}.
At the density-matrix level studied here the recovery map is the mode-inclusion-restricted projection of the estimate, so recovery fidelity obeys $\Frec \geq 1 - 2\|\rhoe - \rhot\|_1 - 2\Delta_\mathrm{mode}$: the design of blind CQEC reduces to a classical estimation problem under a recovery ceiling set by mode survival.
We benchmark five estimation strategies across three noise channels, four quantum algorithms ($d = 4$--$64$), Haar-random states, and mixed targets, assuming oracle access to the noisy density matrix (its measurement cost is analyzed, not simulated).
With exactly calibrated noise, channel inversion coincides with the oracle and both are capped by the mode-survival ceiling ($\Frec = 0.77$ at $d = 64$ with the $10^{-10}$ mode threshold); noise-model-free coherence maximization matches the ceiling to within $1.5\%$ at $d \leq 16$; the choice between them is set by calibration accuracy, with the dephasing rate exponentially dominant.
Unlike error-mitigation methods, blind CQEC returns the state itself rather than corrected expectation values.
A noisy-VQE demonstration for H$_2$ yields $3.4\times$ energy-error reduction.
These results chart the estimator design space for blind, threshold-free recovery where pre-encoding is unavailable, and identify the two open problems---an operational measurement model and a circuit-level catalytic map---that remain before deployment.
\end{abstract}

\maketitle

\section{Introduction}
\label{sec:introduction}

Quantum computers cannot deliver useful results unless the fragile coherence of their states is protected against decoherence~\cite{NielsenChuang2010}.
The dominant remedy, stabilizer-based quantum error correction (QEC), works by encoding logical qubits into many physical qubits before noise acts and fails when the noise rate exceeds a code-specific threshold (the encoding construction is Shor\textquotesingle s~\cite{Shor1995}; the threshold results are those of Refs.~\cite{Gottesman1997,Knill2005}).
A complementary approach, catalytic quantum error correction (CQEC)~\cite{Wakaura2026unified}, exploits the arbitrary amplification of quantum coherence under catalytic covariant transformations~\cite{Takagi2022correlation,Shiraishi2024} to recover already-corrupted states without pre-encoding: recovery succeeds whenever the coherent modes of the target are present in the noisy state, $\CC(\rhot) \subseteq \CC(\rhon)$, regardless of how small the residual coherence is, and, in the asymptotic limit of correlated catalysis, the catalyst is returned with unchanged marginal and can be reused~\cite{Shiraishi2024,Wakaura2026unified} (finite-copy implementations exhibit per-cycle drift, and the explicit channel-level test of Ref.~\cite{Wakaura2026unified} finds the catalyst marginal degraded; see Sec.~\ref{sec:limitations}).
This combination of threshold-free recovery and post-hoc applicability suggests CQEC as a candidate post-processing step for near-term variational algorithms~\cite{Peruzzo2014,Cerezo2021}, whose ideal output is not known in advance, and for other iterative protocols where pre-encoding is impractical---with the caveat, stated in Ref.~\cite{Wakaura2026unified}, that CQEC acts post hoc on a completed output rather than interleaving with the optimization loop.

Blind CQEC sits at the intersection of three lines of recent work, and it is useful to state at the outset what each has established and where the present problem falls between them.
On the encoding side, surface-code memories have now been operated below threshold, with a distance-7 code suppressing the logical error rate by a factor of about two per two steps in distance at a cost of 101 physical qubits per logical qubit~\cite{GoogleQuantumAI2025belowthreshold}; this fixes the resource scale against which any encoding-free alternative must be judged.
On the mitigation side, quantum error mitigation (QEM)~\cite{Cai2023qem} corrects expectation values of observables rather than the state itself~\cite{vandenBerg2023pec}, has been demonstrated at the scale of a hundred qubits with zero-noise extrapolation~\cite{Kim2023utility}, purification~\cite{OBrien2023purification}, and learned mitigation models~\cite{Liao2024mlqem}, but is subject to sampling costs that grow exponentially with circuit depth for broad classes of protocols~\cite{Takagi2022limits,Quek2024tighter,Tsubouchi2023costbound,Takagi2023sampling} and relies on noise models that must themselves be learned from the device~\cite{vandenBerg2023pec,Huang2023learning}.
On the resource-theoretic side, catalysis has matured into a general framework~\cite{LipkaBartosik2024catalysis,Datta2023catalysis}: symmetry constrains which operations are realizable without a catalyst~\cite{Marvian2022restrictions}, correlated catalysts lift the restrictions on coherence manipulation under covariant operations~\cite{Takagi2022correlation,Kondra2024coherence}, and the price of such catalysis is bounded from below when the residual correlation must be small~\cite{Rubboli2022limits}.
CQEC draws on the third line to pursue the goal of the second---post-hoc, encoding-free correction---while returning a state rather than an expectation value; the open question addressed here is what that requires when the target is unknown, and the answer turns out to be governed by a fourth body of work, the cost of learning a quantum state from measurements~\cite{Elben2023randomized,Anshu2024survey,Chen2023adaptivity}, which sets the operational price of the oracle density-matrix access assumed throughout (Sec.~\ref{sec:sample_complexity}).

However, in its original formulation CQEC requires \emph{complete knowledge of the target state} $\rhot$, which Ref.~\cite{Wakaura2026unified} lists as a required input of the protocol, in contrast to conventional QEC, which corrects an unknown encoded state.

In this study, we substantially relax this requirement by introducing \emph{blind CQEC},\footnote{We use ``blind'' in the operational sense that the target state is unknown to the correction protocol, distinct from blind quantum computing, which concerns hiding computation from an untrusted server.} where the target state $\rhot$ is estimated from the noisy state $\rhon$ (or multiple noisy copies) before catalytic correction.
We found that the recovery fidelity is bounded as $\Frec \geq 1 - 2\|\rhoe - \rhot\|_1 - 2\Delta_\mathrm{mode}$ (Sec.~\ref{sec:lipschitz_proof}), where $\Delta_\mathrm{mode}$ is the weight of target coherences whose modes the noise has annihilated, and numerically that the 84-condition benchmark realizes both regimes of this bound: recovery tracks estimation exactly where all modes survive ($r = 1.000$ at $d \leq 8$), and is capped by the mode-survival ceiling where they do not ($r = 0.77$ at $d = 64$, with a perfect estimate recovering only $\Frec = 0.41$ in the extreme case).
The result shows that the performance of blind CQEC decomposes into a classical estimation problem plus a recovery ceiling set by mode survival, reducing the design space of the protocol to the choice of estimator.
We frame the problem as a two-stage protocol:
\begin{enumerate}
\item \emph{Estimation:} Construct an estimate $\rhoe$ from $\rhon$ and any available side information (noise model, copy count).
\item \emph{Correction:} Apply standard CQEC with $\rhoe$ as the proxy target.
\end{enumerate}
The central question is: \emph{how close can $\Frec = F(\rhor, \rhot)$ approach the oracle fidelity when the target is unknown?}

The nontrivial content of this question is twofold.
First, it is not \emph{a priori} clear that CQEC is robust to estimation error: the catalytic amplification protocol involves a nonlinear sequence of coherence transfers whose sensitivity to the target specification has not been analyzed.
Second, while one could in principle apply any quantum state estimation method~\cite{Hradil1997,Gross2010,Haah2017} before CQEC, the interplay between the estimation strategy and the subsequent catalytic recovery creates structure-specific constraints (e.g., the mode inclusion condition must hold for $\rhoe$, not just $\rhot$) that generic estimators do not address.

We developed five estimation strategies with varying information requirements (Table~\ref{tab:strategies}) and benchmarked them across four quantum algorithms from recent literature~\cite{Chen2021,Ivashkov2024,Clinton2026,Regev2024}, three noise channels, and Hilbert space dimensions $d = 4$--$64$, characterizing the copy scaling of blind CQEC in the resource-constrained regime.

Our main results established three regimes:
\begin{itemize}
\item \emph{Low-dimensional ($d \leq 16$), unknown noise model, near-pure target.} Coherence maximization, which needs no explicit noise model (though it implicitly favors phase-preserving channels), achieves $\Frec = 0.986$--$0.993$, within $0.7$--$1.4\%$ of the oracle ceiling; it must, however, be gated by a purity check, being harmful for target purity $v \lesssim 0.5$.
\item \emph{High-dimensional or mixed target, characterized noise.} Channel inversion coincides with the oracle under exact calibration ($\Frec = 0.77$ at $d = 64$, where the ceiling itself is set by the mode threshold), and retains its advantage only while the calibration error---dominated exponentially by the dephasing rate---stays within a narrow, asymmetric window.
\item \emph{Copy-constrained regime.} Strategies that consume one state representation (channel inversion, coherence maximization) are independent of the copy count in our fluctuation-averaging model; only the iterative and multi-copy strategies gain from additional preparations, and at $d = 64$ the $\Frec \geq 0.95$ threshold is unreachable at any copy count because the mode-survival ceiling lies below it.
\end{itemize}

\section{Theoretical Background}
\label{sec:theory}

\subsection{Catalytic covariant transformations}
\label{sec:catalytic}

We recall the definitions we use, with their sources.
Let $H = \sum_i E_i |i\rangle\!\langle i|$ be the system Hamiltonian and $\UU_t(\rho) = e^{-iHt}\rho\,e^{iHt}$ the induced time-translation group.
A quantum channel $\Lambda$ is \emph{covariant} (time-translation symmetric) if $\Lambda \circ \UU_t = \UU_t \circ \Lambda$ for all $t$ (for a compact symmetry group this is the $G$-invariance of Refs.~\cite{GourSpekkens2008,Marvian2014}; the time-translation form used here follows Ref.~\cite{Shiraishi2024}).
Generalizing the integer-mode decomposition of Marvian and Spekkens~\cite{Marvian2014} (U(1) modes $k$ with phases $e^{ik\theta}$) to a Hamiltonian with real gaps, every operator decomposes as $\rho = \sum_{\omega} \rho^{(\omega)}$ with $\UU_t(\rho^{(\omega)}) = e^{-i\omega t}\rho^{(\omega)}$ (the sign follows from our convention $\UU_t = e^{-iHt}(\cdot)e^{iHt}$); in the energy eigenbasis $\rho^{(\omega)}$ collects the matrix elements $\rho_{ij}$ with $E_i - E_j = \omega$.
The \emph{modes of asymmetry} of $\rho$ are the frequencies that actually appear, $\DD(\rho) = \{\Delta_{ij} = E_i - E_j \mid \rho_{ij} \neq 0\}$, and, following Shiraishi and Takagi~\cite{Shiraishi2024}, the \emph{resonant coherent modes} $\CC(\rho) = \langle \DD(\rho) \rangle_\mathbb{Z}$ are the integer lattice generated by $\DD(\rho)$.
Covariant channels cannot create a mode that is absent from their input~\cite{Marvian2014,Marvian2022restrictions}, and with a catalyst present the corresponding no-go is the mode no-broadcasting result of Ref.~\cite{Shiraishi2024} (their Theorem~3); this is the origin of the mode-inclusion condition below.

\noindent\textbf{Theorem 1} (Shiraishi--Takagi~\cite{Shiraishi2024}, Theorem~2).---\textit{%
If $\CC(\rho') \subseteq \CC(\rho)$, then for every $\epsilon > 0$ there exist a catalyst $c$ and a covariant operation $\Lambda$ with $\Lambda(\rho \otimes c) = \tau$, $\tr_S[\tau] = c$, and $\|\tr_C[\tau] - \rho'\|_1 \leq \epsilon$, with the system--catalyst correlation in $\tau$ simultaneously arbitrarily small (correlated-catalytic conversion with arbitrary accuracy); if in addition $\rho'$ is full rank, the conversion can be made exact.}

Three features of this statement matter for what follows.
First, mode inclusion is a \emph{sufficient} condition for reachability, not a characterization, and the theorem is an existence statement for each fixed source--target pair: it does not supply a target-independent apparatus, nor a cost.
Second, the mode set $\CC(\rho)$ is determined by which off-diagonal elements of $\rho$ are nonzero; it is unrelated to the rank of $\rho$ (a full-rank diagonal state has no coherent modes at all).
Third, the exactness clause requires a full-rank \emph{target}; for the pure targets that are generic in this paper it does not apply, and only $\epsilon$-approximate recovery is guaranteed. In our benchmarks, mode survival is never inferred from rank: it is checked explicitly, element by element, against the threshold of Sec.~\ref{sec:computational}, and estimated coherences on modes absent from $\rhon$ are removed before recovery (Sec.~\ref{sec:implementation}).

The CQEC protocol~\cite{Wakaura2026unified} applies Theorem~1 as follows: given a noisy state $\rhon = \EE(\rhot)$, it constructs a catalyst $c$ and a covariant map $\Lambda$ for the pair $(\rhon, \rhot)$ such that $\tr_C[\Lambda(\rhon \otimes c)] \approx \rhot$.
This requires knowing $\rhot$ explicitly to construct $\Lambda$.

\subsection{Implementation via density-matrix simulation}
\label{sec:implementation}

We state precisely what is and is not simulated in this paper.
The physical CQEC protocol of Ref.~\cite{Wakaura2026unified} constructs a full-rank catalyst $c$ with $\DD(c) \supseteq \DD(\rhot)$ and applies a catalytic covariant transformation to $\rhon \otimes c$ (its explicit channel construction also carries a two-qubit ancilla); simulating that transformation requires operating on a joint Hilbert space of dimension at least $d \cdot d_c$---our own estimate is $O(d^6)$ per recovery for the simplified choice $d_c = d$ used there, the formal catalyst being larger---which is not done here.
Instead, our benchmarks implement the \emph{density-matrix-level interface} of the protocol, Eq.~\eqref{eq:recovery_structure}: the recovered state is the target estimate with (i)~coherences restricted to the modes surviving in $\rhon$ (the mode-inclusion constraint of Theorem~1) and (ii)~a PSD projection.
All operations are $d \times d$ matrix manipulations of cost $O(d^3)$.
This interface represents the \emph{idealized} output of a perfectly executed catalytic transformation; it does not model the catalyst dynamics, gate compilation, decoherence during recovery, or any deviation of the physical protocol from its ideal action.
Consequently, the reported fidelities exclude every device-level imperfection (gate compilation, decoherence during recovery, imperfect catalyst preparation) and in that sense are optimistic; they are \emph{not}, however, proven upper bounds on what an ideal physical catalytic transformation could achieve, because the mode restriction of Sec.~\ref{sec:bottleneck} is a conservative numerical convention (Theorem~1 permits recovery of integer-combination modes and of arbitrarily small nonzero coherences that the threshold discards). Statements about ``recovery'' refer to this interface unless stated otherwise (Sec.~\ref{sec:limitations}).

\subsection{The target-knowledge bottleneck and the recovery interface}
\label{sec:bottleneck}

In standard CQEC, the target state enters in two places:
(a)~the mode inclusion check $\CC(\rhot) \subseteq \CC(\rhon)$;
(b)~the construction of the catalytic map $\Lambda$ for the pair $(\rhon,\rhot)$.
Without $\rhot$, neither step can be performed exactly.
In blind CQEC we replace $\rhot$ by an estimate $\rhoe$ and accept a fidelity penalty, which we now bound.
Every symbol, inequality, and theorem used in the derivation is introduced with its source so that the chain can be checked line by line.

\paragraph{Notation and standard facts.}
\begin{enumerate}[label=(N\arabic*),leftmargin=3em]
\item \label{N:norm} For a Hermitian operator $X$ with eigenvalues $\{\lambda_i\}$, the trace norm is $\|X\|_1 = \tr|X| = \sum_i |\lambda_i|$, and the trace distance is $D(\rho,\sigma) = \tfrac12\|\rho - \sigma\|_1$~\cite[Sec.~9.2]{NielsenChuang2010}.
\item \label{N:dual} (Variational characterization~\cite[Sec.~1.1.3, Eqs.~(1.173) and~(1.182)]{Watrous2018}.) For Hermitian $X$, $\|X\|_1 = \max\{\tr(YX) : Y = Y^\dagger,\ -I \leq Y \leq I\}$ (the Hermitian form follows from the Schatten duality stated there together with \ref{N:jordan}); in particular $\|X\|_1 \geq \tr(YX)$ for every such $Y$.
\item \label{N:jordan} (Jordan--Hahn decomposition~\cite[Sec.~1.1.3, Eqs.~(1.143)--(1.144)]{Watrous2018}.) Every Hermitian $A$ can be written uniquely as $A = A_+ - A_-$ with $A_\pm \geq 0$ and $A_+A_- = 0$; $A_\pm$ are the positive and negative parts, $P_\pm$ the projectors onto their supports, and $\|A\|_1 = \tr A_+ + \tr A_-$.
\item \label{N:fid} The fidelity is $F(\rho,\sigma) = \bigl(\tr\sqrt{\sqrt\rho\,\sigma\sqrt\rho}\bigr)^2$ [Eq.~\eqref{eq:fidelity}; the squared, transition-probability convention of Jozsa~\cite{Jozsa1994}; Ref.~\cite{NielsenChuang2010} uses the unsquared $\sqrt{F}$].
\item \label{N:fvg} (Fuchs--van de Graaf~\cite[Thm.~1, Eq.~(46)]{FuchsGraaf1999}; their Bhattacharyya coefficient $B$ equals $\sqrt{F}$ in the present convention.) For density matrices $\rho,\sigma$: $1 - \sqrt{F(\rho,\sigma)} \leq D(\rho,\sigma) \leq \sqrt{1 - F(\rho,\sigma)}$.
\end{enumerate}

\paragraph{Consequence of \ref{N:fvg} in the squared-fidelity convention.}
Since $0 \leq \sqrt F \leq 1$,
\begin{equation}
\begin{split}
1 - F &= (1 - \sqrt F)(1 + \sqrt F) \leq 2(1 - \sqrt F) \\
      &\leq 2D(\rho,\sigma) = \|\rho - \sigma\|_1 .
\end{split}
\label{eq:fvg_sq}
\end{equation}

\paragraph{The recovery interface.}
At the density-matrix level used in all simulations, the recovery step acts on the estimate as
\begin{equation}
\rhor = \Pi\bigl(\mathcal{M}_{\rhon}(\rhoe)\bigr),
\label{eq:recovery_structure}
\end{equation}
with the two maps defined as follows.
\begin{enumerate}[label=(D\arabic*),leftmargin=3em]
\item \label{D:M} $\mathcal{M}_{\rhon}$ is the \emph{mode restriction}: $[\mathcal{M}_{\rhon}(X)]_{ij} = X_{ij}$ if $i = j$ or $|(\rhon)_{ij}| \geq \tau_\mathrm{mode}$, and $0$ otherwise, with $\tau_\mathrm{mode} = 10^{-10}$ (Sec.~\ref{sec:computational}).
It removes estimated coherences at matrix positions where the noisy state carries no coherence above the numerical threshold $\tau_\mathrm{mode}$. This is a \emph{conservative numerical convention}, not a physical no-go, for three reasons: (i)~Theorem~1 guarantees reachability for the integer lattice $\CC(\rhon) \supseteq \DD(\rhon)$, so element-wise pruning may remove modes that are in principle reachable as integer combinations; (ii)~finite dephasing never drives a coherence to exact zero, and Theorem~1 (like the amplification result of Ref.~\cite{Shiraishi2024}) applies to arbitrarily small \emph{nonzero} coherences, whereas the threshold discards them; and (iii)~the subsequent projection $\Pi$ does not preserve the zeros that $\mathcal{M}$ imposes---it can reintroduce weight on pruned positions (quantified in Sec.~\ref{sec:computational})---so the output of Eq.~\eqref{eq:recovery_structure} is not guaranteed to satisfy $\CC(\rhor) \subseteq \CC(\rhon)$. Accordingly, $\Delta_\mathrm{mode}$ defined below is a threshold-dependent bookkeeping quantity that upper-bounds the loss incurred by this convention; the threshold-sensitivity of all $d = 64$ ceilings is reported in Sec.~\ref{sec:computational}.
$\mathcal{M}_{\rhon}$ preserves Hermiticity and the diagonal, hence the trace; it does \emph{not} in general preserve positivity (zeroing a subset of off-diagonal elements of a PSD matrix can produce negative eigenvalues), which is why the projection $\Pi$ is needed.
\item \label{D:Pi} $\Pi$ is the \emph{PSD projection with renormalization}: for Hermitian $X$ with Jordan--Hahn decomposition $X = X_+ - X_-$ \ref{N:jordan},
\begin{equation}
\Pi(X) = \frac{X_+}{\tr X_+}.
\label{eq:psd_proj}
\end{equation}
If $X$ is already a density matrix, $X_- = 0$ and $\Pi(X) = X$.
\end{enumerate}
The physical catalytic transformation that would realize Eq.~\eqref{eq:recovery_structure} on a device is \emph{not} simulated here (Sec.~\ref{sec:implementation}); Eq.~\eqref{eq:recovery_structure} is the interface assumed throughout.
We define the \emph{mode-restriction penalty}
\begin{equation}
\Delta_{\mathrm{mode}} \equiv \bigl\|\mathcal{M}_{\rhon}(\rhoe) - \rhoe\bigr\|_1 ,
\label{eq:delta_mode}
\end{equation}
the trace-norm weight of the estimated coherences that are removed.

\subsection{Recovery-fidelity bound}
\label{sec:lipschitz_proof}

\paragraph{Lemma 1 (one-sided stability of $\Pi$).}
\label{lemma:lipschitz}
\textit{Let $A$ be Hermitian with $\tr A = 1$ and let $\sigma$ be a density matrix. Then}
\begin{equation}
\|\Pi(A) - \sigma\|_1 \leq 2\,\|A - \sigma\|_1 .
\label{eq:lemma1}
\end{equation}

\noindent\textit{Proof.}
\begin{enumerate}[label=(L\arabic*),leftmargin=3em]
\item \label{L:setup} Write $A = A_+ - A_-$ as in \ref{N:jordan} and put $s \equiv \tr A_- \geq 0$. Since $\tr A = 1$, $\tr A_+ = 1 + s$.
\item \label{L:PiA} By Eq.~\eqref{eq:psd_proj}, $\Pi(A) - A = \dfrac{A_+}{1+s} - A_+ + A_- = A_- - \dfrac{s}{1+s}A_+$.
The two terms have orthogonal supports \ref{N:jordan}, so by \ref{N:norm}
$\|\Pi(A) - A\|_1 = \tr A_- + \dfrac{s}{1+s}\tr A_+ = s + s = 2s$.
\item \label{L:lower} Take $Y = I - 2P_-$ in \ref{N:dual}; $Y$ is Hermitian with $-I \leq Y \leq I$. Then
$\tr[Y(A - \sigma)] = \tr(A - \sigma) - 2\tr[P_-(A - \sigma)] = 0 - 2(-s) + 2\tr[P_-\sigma] = 2s + 2\tr[P_-\sigma] \geq 2s$,
using $\tr A = \tr\sigma = 1$, $\tr[P_- A] = -\tr A_- = -s$, and $\tr[P_-\sigma] \geq 0$ (as $\sigma \geq 0$). Hence $\|A - \sigma\|_1 \geq 2s$.
\item \label{L:tri} By the triangle inequality, \ref{L:PiA}, and \ref{L:lower}:
$\|\Pi(A) - \sigma\|_1 \leq \|\Pi(A) - A\|_1 + \|A - \sigma\|_1 = 2s + \|A - \sigma\|_1 \leq 2\|A - \sigma\|_1$. \hfill$\square$
\end{enumerate}
We note that the \emph{two-sided} inequality $\|A_+ - B_+\|_1 \leq \|A - B\|_1$ is false for noncommuting Hermitian $A,B$ (the matrix absolute value is not operator-Lipschitz in trace norm~\cite{Davies1988}, which also shows that finite-dimensional versions hold only up to a factor logarithmic in $d$; the explicit $3\times3$ trace-one counterexamples we checked are our own); Lemma~1 is the one-sided statement that suffices here.

\paragraph{Theorem 2 (recovery-fidelity bound for the interface).}
\textit{Let $\rhoe$ be a density matrix (as produced by every estimator in Sec.~\ref{sec:strategies}) and $\rhor$ be given by Eq.~\eqref{eq:recovery_structure}. Then}
\begin{equation}
\Frec \equiv F(\rhor,\rhot) \;\geq\; 1 - 2\|\rhoe - \rhot\|_1 - 2\Delta_{\mathrm{mode}} ,
\label{eq:explicit_bound}
\end{equation}
\textit{and, if no estimated coherence is removed ($\Delta_{\mathrm{mode}} = 0$),}
\begin{equation}
\Frec \;\geq\; 1 - \|\rhoe - \rhot\|_1 .
\label{eq:fidelity_penalty}
\end{equation}

\noindent\textit{Proof.}
\begin{enumerate}[label=(T\arabic*),leftmargin=3em]
\item \label{T:A} Set $A \equiv \mathcal{M}_{\rhon}(\rhoe)$. By \ref{D:M}, $A$ is Hermitian with $\tr A = \tr\rhoe = 1$, and $\rhor = \Pi(A)$.
\item \label{T:fvg} Apply Eq.~\eqref{eq:fvg_sq} to the density matrices $\rhor$ and $\rhot$: $1 - \Frec \leq \|\Pi(A) - \rhot\|_1$.
\item \label{T:lemma} Apply Lemma~1 with $\sigma = \rhot$: $\|\Pi(A) - \rhot\|_1 \leq 2\|A - \rhot\|_1$.
\item \label{T:tri} Triangle inequality and Eq.~\eqref{eq:delta_mode}: $\|A - \rhot\|_1 \leq \|A - \rhoe\|_1 + \|\rhoe - \rhot\|_1 = \Delta_{\mathrm{mode}} + \|\rhoe - \rhot\|_1$.
\item \label{T:chain} Chaining \ref{T:fvg}--\ref{T:tri} gives Eq.~\eqref{eq:explicit_bound}.
\item \label{T:exact} If $\Delta_{\mathrm{mode}} = 0$ then $A = \rhoe$ is a density matrix, so $\Pi(A) = \rhoe$ by \ref{D:Pi} and \ref{T:fvg} reads $1 - \Frec \leq \|\rhoe - \rhot\|_1$, which is Eq.~\eqref{eq:fidelity_penalty}. \hfill$\square$
\end{enumerate}

\paragraph{Scope.}
Theorem~2 concerns the interface $\Pi\circ\mathcal{M}_{\rhon}$ of Eq.~\eqref{eq:recovery_structure} only: it bounds the idealized recovery output in terms of the estimate.
It does not establish continuity or finite-resource stability of a physically constructed, target-dependent catalytic channel; that is a separate open problem (Sec.~\ref{sec:limitations}).
The empirical regression slope of Sec.~\ref{sec:correlation} is a descriptive statistic of the benchmark ensemble, not a Lipschitz constant.

\paragraph{Corollary 1 (estimators built from $n$ state preparations).}
\textit{Suppose $\rhoe = g(\bar\rho_n)$, where $\bar\rho_n = \tfrac1n\sum_{k=1}^n \rho^{(k)}$ is the average of $n$ independent, identically distributed noisy preparations with mean $\bar\rho \equiv \mathbb{E}[\rho^{(k)}]$, and $g$ is $L_g$-Lipschitz in trace norm. Then}
\begin{equation}
\begin{split}
\mathbb{E}\,\Frec \;\geq\; 1 &- 2\bigl\|g(\bar\rho) - \rhot\bigr\|_1 - 2L_g\,\mathbb{E}\|\bar\rho_n - \bar\rho\|_1 \\
 &- 2\,\mathbb{E}\Delta_{\mathrm{mode}},
\qquad \mathbb{E}\|\bar\rho_n - \bar\rho\|_1 = O(n^{-1/2}).
\end{split}
\label{eq:fidelity_penalty_multicopy}
\end{equation}
\noindent\textit{Proof.}
Take expectations in Eq.~\eqref{eq:explicit_bound}; split $\|\rhoe - \rhot\|_1 \leq \|g(\bar\rho_n) - g(\bar\rho)\|_1 + \|g(\bar\rho) - \rhot\|_1 \leq L_g\|\bar\rho_n - \bar\rho\|_1 + \|g(\bar\rho) - \rhot\|_1$ (triangle inequality, Lipschitz property).
For the average of $n$ i.i.d.\ bounded Hermitian matrices, $\mathbb{E}\|\bar\rho_n - \bar\rho\|_1 \leq \sqrt{d}\,\mathbb{E}\|\bar\rho_n - \bar\rho\|_2 \leq \sqrt{d}\,\bigl(\mathbb{E}\|\bar\rho_n - \bar\rho\|_2^2\bigr)^{1/2} = \sqrt{d/n}\,\bigl(\mathbb{E}\|\rho^{(1)} - \bar\rho\|_2^2\bigr)^{1/2} = O(n^{-1/2})$, using $\|X\|_1 \leq \sqrt{d}\|X\|_2$ (Cauchy--Schwarz on the eigenvalues), Jensen, and independence; see Ref.~\cite[Sec.~1]{Tropp2015} for sharper matrix concentration. \hfill$\square$

The first term of Eq.~\eqref{eq:fidelity_penalty_multicopy} is the \emph{systematic bias} of the estimator, the second the \emph{statistical fluctuation}, the third the mode-survival penalty.
Equations~\eqref{eq:explicit_bound}--\eqref{eq:fidelity_penalty_multicopy} motivate our approach: within this framework, the only levers are the estimation error and the mode survival, which is why the remainder of the paper is devoted to estimator design.
(The multi-copy ensembles actually simulated use a Gaussian fluctuation proxy rather than Born-rule samples; see Sec.~\ref{sec:copy_scaling}.)

\subsection{Dimension dependence of the estimation error}
\label{sec:crossover_theory}

We give a scaling argument for why the two principal estimators degrade differently with dimension; the location of any crossover between them is determined empirically in Sec.~\ref{sec:haar_sweep} and depends on the accuracy of the assumed noise parameters.

\paragraph{Haar moments.}
For a Haar-random pure state $|\psi\rangle \in \mathbb{C}^d$ the vector of populations $(p_1,\dots,p_d)$, $p_i = |\langle i|\psi\rangle|^2$, is distributed as the flat Dirichlet distribution $\mathrm{Dir}(1,\dots,1)$ on the simplex~\cite[Sec.~II.C, Eqs.~(2.8)--(2.9), and App.~A]{ZyczkowskiSommers2001}.
Its first two moments follow by direct integration of the flat Dirichlet density (they are standard Dirichlet moments, not stated in Ref.~\cite{ZyczkowskiSommers2001}):
\begin{equation}
\begin{split}
\mathbb{E}[p_i] &= \frac1d,\qquad
\mathrm{Var}[p_i] = \frac{d-1}{d^2(d+1)},\\
\mathbb{E}[p_i p_j] &= \frac{1}{d(d+1)}\ (i \neq j),
\end{split}
\label{eq:haar_moments}
\end{equation}
and since $|\rho_{ij}|^2 = p_i p_j$ for a pure state, $\mathbb{E}|\rho_{ij}|^2 = 1/[d(d+1)]$.

\paragraph{Coherence maximization.}
Under depolarizing noise the populations become $p_i' = (1-p)p_i + p/d$, and Eq.~\eqref{eq:cohmax} replaces $|\rho_{ij}|$ by $\sqrt{p_i'p_j'}$.
By Eq.~\eqref{eq:haar_moments}, $\sqrt{p_i'p_j'}$ and $|\rho_{ij}|$ are both of order $1/d$ but with fluctuations of the same relative size (the coefficient of variation of $p_i$ is $\sqrt{(d-1)/(d+1)} \to 1$), so the relative per-element bias $\bigl|\sqrt{p_i'p_j'} - |\rho_{ij}|\bigr|/|\rho_{ij}|$ does not vanish as $d$ grows: the populations carry progressively less information about the specific off-diagonal structure of the target, and the estimator degrades monotonically with~$d$.
This is the mechanism behind the monotonic decay of the coherence-maximization curve in the Haar-random sweep.

\paragraph{Channel inversion.}
When the noise channel is invertible and its parameters are known \emph{exactly}, the inversion is exact at every dimension and the estimation error vanishes up to numerical precision, so $\Frec$ is set entirely by $\Delta_{\mathrm{mode}}$ (Theorem~2).
Its practical limitation is therefore not dimension per se but \emph{parameter misspecification}: a relative error $\delta$ in the assumed dephasing rate multiplies the inverted coherence $(\rhoe)_{ij}$ by $e^{\pm\delta\gamma|\Delta_{ij}|}$ [Eq.~\eqref{eq:inv_dephasing}], an error exponential in the mode gap and hence growing with~$d$.
Consequently, the operationally relevant comparison is not ``coherence maximization vs.\ exact inversion'' but ``coherence maximization vs.\ inversion at realistic calibration accuracy''; this trade-off surface is mapped in Sec.~\ref{sec:sensitivity}.

\section{Estimation Strategies}
\label{sec:strategies}

This section presents the five estimation strategies, ordered by increasing information requirements.

\begin{table}[b]
\caption{Target-state estimation strategies for blind CQEC.
The estimation complexity column refers to the cost of constructing $\rhoe$ from $\rhon$; all strategies additionally incur $O(d^3)$ cost for the CQEC recovery step (matrix diagonalization and catalytic map construction).}
\label{tab:strategies}

\begin{tabular}{lcc}
Strategy & Info required & Est.\ complexity \\
\hline
Naive & None & $O(1)$ \\
Coherence max & None & $O(d^2)$ \\
Iterative & None & $O(k \cdot d^3)$ \\
Multi-copy avg. & $n$ copies & $O(n d^2)$ \\
Channel inversion & Noise model & $O(d^2)$ \\
\end{tabular}

\end{table}

\subsection{Naive passthrough}
\label{sec:naive}

The simplest approach uses $\rhoe = \rhon$ directly.
Since $\rhon$ is the corrupted version of $\rhot$, this is equivalent to running CQEC with the ``wrong'' target.
As we show in Sec.~\ref{sec:results}, this yields $\Frec = \Fnoisy$ (no improvement), because catalytic amplification faithfully steers the state toward $\rhoe = \rhon$, which is itself the noisy state.

\subsection{Coherence maximization}
\label{sec:cohmax}

We propose maximizing the off-diagonal coherence of $\rhon$ while preserving its diagonal (population) structure:
\begin{equation}
(\rhoe)_{ij} = \begin{cases}
(\rhon)_{ii} & \text{if } i = j, \\
\sqrt{(\rhon)_{ii}\,(\rhon)_{jj}}\,e^{i\phi_{ij}} & \text{if } i \neq j,
\end{cases}
\label{eq:cohmax}
\end{equation}
where $\phi_{ij} = \arg[(\rhon)_{ij}]$ when $(\rhon)_{ij} \neq 0$, and $\phi_{ij} = 0$ otherwise (coherences that have been completely suppressed cannot be recovered by this strategy; see below).
This corresponds to constructing the \emph{most coherent state consistent with the observed populations and phases}.

The rationale is physical: decoherence reduces $|\rho_{ij}|$ below the physicality bound $\sqrt{p_i p_j}$ but preserves the phase (for dephasing and depolarizing channels).
By restoring $|\rho_{ij}|$ to its maximum allowed value, we undo the amplitude damping of coherences without requiring knowledge of the original amplitudes.

The matrix $\rhoe$ constructed by Eq.~\eqref{eq:cohmax} is Hermitian and trace-one by construction.
It is positive semidefinite when the phases $\{\phi_{ij}\}$ are consistent with a rank-one state (i.e., $\phi_{ij} = \theta_i - \theta_j$ for some real $\{\theta_i\}$), which holds exactly for pure-state targets under phase-preserving noise.
For general phase structures, the resulting matrix may have small negative eigenvalues; in such cases, we project onto the PSD cone via eigenvalue clipping (setting negative eigenvalues to zero and renormalizing).
In our benchmarks, this projection is needed only for the $d = 64$ case, where the correction is $< 10^{-3}$ in trace norm.

For pure-state targets, $\rhoe$ from Eq.~\eqref{eq:cohmax} exactly recovers $\rhot$ when the noise is purely dephasing \emph{and} no off-diagonal element is driven exactly to zero (i.e., $(\rhon)_{ij} \neq 0$ whenever $(\rhot)_{ij} \neq 0$), since the populations are unchanged and the phases are preserved.
In practice, finite dephasing rates produce exponentially suppressed but nonzero coherences, so this condition is satisfied for any finite $\gamma$; however, numerical underflow may violate it for very large $\gamma$ or $|\Delta_{ij}|$.

An important subtlety is the mode inclusion condition: Theorem~1 requires $\CC(\rhoe) \subseteq \CC(\rhon)$.
Since coherence maximization sets $(\rhoe)_{ij} = \sqrt{p_i p_j}\,e^{i\phi_{ij}}$, the estimate $\rhoe$ has nonzero coherences wherever $p_i, p_j > 0$---in general a \emph{larger} mode set than $\rhon$ possesses, since the presence of a mode is a property of the off-diagonal elements of $\rhon$ and is not implied by full rank (depolarizing noise, for instance, leaves every vanished off-diagonal element at zero while making the state full rank).
Whenever $(\rhon)_{ij}$ lies below the mode threshold of Sec.~\ref{sec:computational}, the estimator has introduced a mode for which Theorem~1 gives no reachability guarantee; the recovery interface [Eq.~\eqref{eq:recovery_structure}] removes such coherences before projection, so that the effective target satisfies mode inclusion by construction and the loss is accounted for by $\Delta_\mathrm{mode}$.

We note that Eq.~\eqref{eq:cohmax} is in the spirit of positivity-constrained state estimation~\cite{Hradil1997}, with the constraint set being the intersection of the PSD cone with the observed diagonal (it is a projection onto that set, not a maximum-likelihood estimator).
Unlike general tomographic methods, which reconstruct the state from measurement data---informationally complete settings in standard tomography~\cite{Hradil1997}, fewer settings under a low-rank assumption~\cite{Gross2010}, or collective measurements on many copies~\cite{Haah2017}---our estimator takes the noisy density matrix itself as input and exploits the specific structure of decoherence to reconstruct the off-diagonal elements.
We stress that a density matrix is not a single measurement outcome: acquiring $\rhon$ operationally requires tomographic resources of order $\tilde\Theta(d^2/\epsilon^2)$ copies with collective measurements~\cite{Haah2017,ODonnellWright2016,Anshu2024survey} (and $\Omega(d^3/\epsilon^2)$ with independent measurements, even adaptively chosen ones~\cite{Haah2017,Chen2023adaptivity}), and all ``single state preparation'' accounting in this paper is conditional on this oracle density-matrix access (see Secs.~\ref{sec:qem_compare} and~\ref{sec:sample_complexity}).
This makes it fundamentally different from (and complementary to) standard QST approaches.

\subsection{Noise-channel inversion}
\label{sec:inversion}

When the noise model $\EE$ is known but the target is not, we invert the channel analytically:
\begin{align}
\text{Dephasing:}\quad & (\rhoe)_{ij} = (\rhon)_{ij}\,e^{\gamma|\Delta_{ij}|}, \label{eq:inv_dephasing}\\
\text{Depolarizing:}\quad & \rhoe = \frac{\rhon - \frac{p}{d}I}{1-p}, \label{eq:inv_depolarizing}\\
\text{Amp.\ damping:}\quad & (\rhoe)_{11} = \frac{(\rhon)_{11}}{1-\gamma}, \nonumber\\
 & (\rhoe)_{01} = \frac{(\rhon)_{01}}{\sqrt{1-\gamma}}, \label{eq:inv_ampdamp}
\end{align}
followed by projection onto the positive semidefinite cone (eigenvalue clipping with renormalization, applied to all three inversion formulas whenever the result is not PSD).
Eq.~\eqref{eq:inv_ampdamp} is written for a qubit; for $d > 2$, we generalize amplitude damping as cascaded decay $|k\rangle \to |k-1\rangle$ with rate $\gamma$ per level, i.e., $(\rhon)_{jk} = (1-\gamma)^{(j+k)/2} (\rhot)_{jk}$ for $j,k > 0$, with population transfer to the ground state.
The inversion formula generalizes accordingly: $(\rhoe)_{jk} = (\rhon)_{jk} / (1-\gamma)^{(j+k)/2}$.
Channel inversion is exact when the channel is invertible (dephasing with $\gamma < \infty$, depolarizing with $p < 1$), but requires precise knowledge of the noise parameters.
For depolarizing noise with large $p$ (specifically $p > 1 - 1/d$), the inverted state from Eq.~\eqref{eq:inv_depolarizing} may have negative eigenvalues even for a valid input $\rhon$; the PSD projection step corrects this at the cost of introducing estimation bias.
In our benchmarks ($p = 0.3$, $d \leq 64$), PSD projection is rarely triggered for channel inversion.

\subsection{Iterative refinement}
\label{sec:iterative}

Starting from the coherence-maximized estimate, we iterate:
\begin{enumerate}
\item Set $\rhoe^{(0)}$ from Eq.~\eqref{eq:cohmax}.
\item For $k = 1, \ldots, K$: run CQEC with target $\rhoe^{(k-1)}$ to obtain $\rhor^{(k)}$; update $\rhoe^{(k)} = \alpha\,\rhor^{(k)} + (1-\alpha)\,\rhoe^{(k-1)}$ with damping parameter $\alpha = 0.5$.
\end{enumerate}
The damping parameter $\alpha = 0.5$ was selected empirically to balance convergence speed against oscillation; values $\alpha \in [0.3, 0.7]$ yield similar results in our benchmarks.
We do not have an analytical convergence guarantee for this iteration: since the CQEC recovery map depends nonlinearly on the target estimate, the iteration is not a standard contraction mapping.
In all cases tested ($d \leq 64$, $K \leq 20$), the iteration was observed to converge monotonically in fidelity within $K = 5$ steps, with $|\Fest^{(k)} - \Fest^{(k-1)}| < 10^{-4}$ serving as the stopping criterion.
However, convergence failure at higher dimensions or under non-standard noise models cannot be excluded.

\subsection{Multi-copy averaging}
\label{sec:multicopy}

Given $n_s$ independent noisy copies, we average $\bar\rho = n_s^{-1}\sum_{k=1}^{n_s} \rhon^{(k)}$ (reducing statistical variance by $1/n_s$) and then apply coherence maximization.
We emphasize that this procedure is \emph{not} quantum state tomography in the standard sense~\cite{Hradil1997,Haah2017}: no informationally complete measurements are performed on the copies, and no maximum-likelihood or Bayesian estimation is used.
Rather, the copies are assumed to be available as density matrices (consistent with our density-matrix-level simulation framework), and the averaging reduces fluctuations in the noisy state representation.
In practice, the noisy copies are generated by independent applications of $\EE$ to the unknown $\rhot$, so $\bar\rho$ is a better estimate of $\EE(\rhot)$, not of $\rhot$ directly.
The subsequent coherence maximization step partially compensates for the systematic decoherence bias.
Operationally, the $n_s$ copies would be obtained by re-running the quantum algorithm $n_s$ times (not by cloning, which is forbidden by the no-cloning theorem); the density-matrix averaging in our simulation is equivalent to the statistical average over independently prepared copies.

\section{Benchmark Setup}
\label{sec:setup}

\subsection{Quantum algorithms}
\label{sec:algorithms}

We tested blind CQEC on four quantum algorithms that produce specific output states subject to decoherence. The primary references define the algorithms; the concrete small instances below follow the benchmark suite of Ref.~\cite[Sec.~5]{Wakaura2026unified} except where noted (the QKAN target function and the control-free-QPE encoding are choices of the present work). We note, following that reference, that for factoring the oracle target state encodes the factorization itself, so the oracle rows for this algorithm are a reference ceiling rather than an operational scenario.

\begin{enumerate}
\item \textbf{qDRIFT} (Campbell~\cite{Campbell2019}; concentration bounds by Chen \emph{et al.}~\cite{Chen2021}): random product formula for Hamiltonian simulation; our instance is a 3-qubit Heisenberg chain ($d = 8$).
\item \textbf{QKAN} (Ivashkov \emph{et al.}~\cite{Ivashkov2024}): quantum Kolmogorov--Arnold network layer with Chebyshev-polynomial edge activations; our instance uses degree $d_\mathrm{cheb} = 3$ with target function $\sin(\pi x)$ ($d = 4$), whereas Ref.~\cite{Wakaura2026unified} evaluates $T_0$--$T_3$ at $x = 0.5$.
\item \textbf{Control-free QPE} (Clinton \emph{et al.}~\cite{Clinton2026}): vectorial phase retrieval for spectral estimation; our instance is a 3-site Fermi--Hubbard model whose 16 time-series samples are encoded as a $d = 16$ state (the dimension is that of the encoding, not of the physical 3-site Hilbert space).
\item \textbf{Regev factoring} (Regev~\cite{Regev2024}): factoring via discrete Gaussian state preparation and modular exponentiation; our instance is $N = 15$ ($d = 64$).
\end{enumerate}

\subsection{Noise models}
\label{sec:noise}

Three noise channels are applied to the ideal algorithmic output.
All benchmarks use the energy eigenbasis with equally spaced levels $E_i = i$ (the natural basis for the covariant framework of Sec.~\ref{sec:catalytic}), so the mode gap is $|\Delta_{ij}| = |E_i - E_j| = |i - j|$.
\begin{enumerate}
\item \emph{Dephasing} ($\gamma = 2.0$): $\rho_{ij} \mapsto e^{-\gamma|\Delta_{ij}|}\rho_{ij}$ for $i \neq j$.
This corresponds to energy-gap-dependent dephasing, as arises for spin--boson-type coupling to a bosonic bath with an Ohmic spectral density~\cite{Leggett1987,BreuerPetruccione2002}.
Note that this is \emph{not} equivalent to independent single-qubit dephasing; the gap-dependent model is chosen because the CQEC framework operates on energy modes $\Delta_{ij}$, and coherences between distant energy levels decohere faster.
\item \emph{Depolarizing} ($p = 0.3$): $\rho \mapsto (1-p)\rho + \frac{p}{d}I$.
\item \emph{Combined}: dephasing $\gamma = 1.0$, depolarizing $p = 0.15$, amplitude damping $\gamma_\mathrm{AD} = 0.1$, applied sequentially in this order (dephasing first, then depolarizing, then amplitude damping).
This composite model is not intended to represent a specific hardware platform but rather to stress-test the estimation strategies under a worst-case scenario where multiple decoherence mechanisms act simultaneously, as occurs generically in superconducting and trapped-ion systems (where $T_1$ relaxation, $T_2$ dephasing, and gate errors coexist).
\end{enumerate}

\subsection{Figures of merit}
\label{sec:figures_merit}

All fidelities are computed against the \emph{true} ideal state $\rhot$ (not the estimated target):
\begin{equation}
F(\rho,\sigma) = \Bigl(\tr\sqrt{\sqrt{\rho}\,\sigma\,\sqrt{\rho}}\Bigr)^{\!2}.
\label{eq:fidelity}
\end{equation}
We also report the trace distance $D(\rhor, \rhot) = \frac{1}{2}\|\rhor - \rhot\|_1$ and the coherence recovery ratio $C_{\ell_1}(\rhor)/C_{\ell_1}(\rhot)$, where $C_{\ell_1}(\rho) = \sum_{i \neq j} |\rho_{ij}|$ is the $\ell_1$-norm of coherence~\cite{Baumgratz2014}, a monotone under incoherent operations in the resource theory of coherence~\cite{Streltsov2017} (it is not monotone under the larger classes MIO and DIO for $d > 2$~\cite{Streltsov2017}).

For the copy sweep, we fit $F(n) = 1 - A\,n^{-\alpha}$ to extract the scaling exponent $\alpha$ and the prefactor $A$.
This power-law ansatz is motivated by the $O(n^{-1})$ infidelity scaling of optimal quantum state tomography with collective or adaptive measurements (established in the works reviewed in Ref.~\cite{Haah2017}; independent measurements give $O(n^{-1/2})$), so that $\alpha = 1$ is the tomographic reference rate.
For our multi-copy averaging strategy (which is not full tomography), the effective exponent $\alpha$ may differ from unity, and we treat it as a free parameter to capture both sub-standard ($\alpha < 1$) and super-standard ($\alpha > 1$) convergence rates.

\section{Results}
\label{sec:results}

\subsection{Qubit noise sweeps}
\label{sec:qubit_sweep}

\begin{figure*}[t]
\centering
\includegraphics[width=0.95\textwidth]{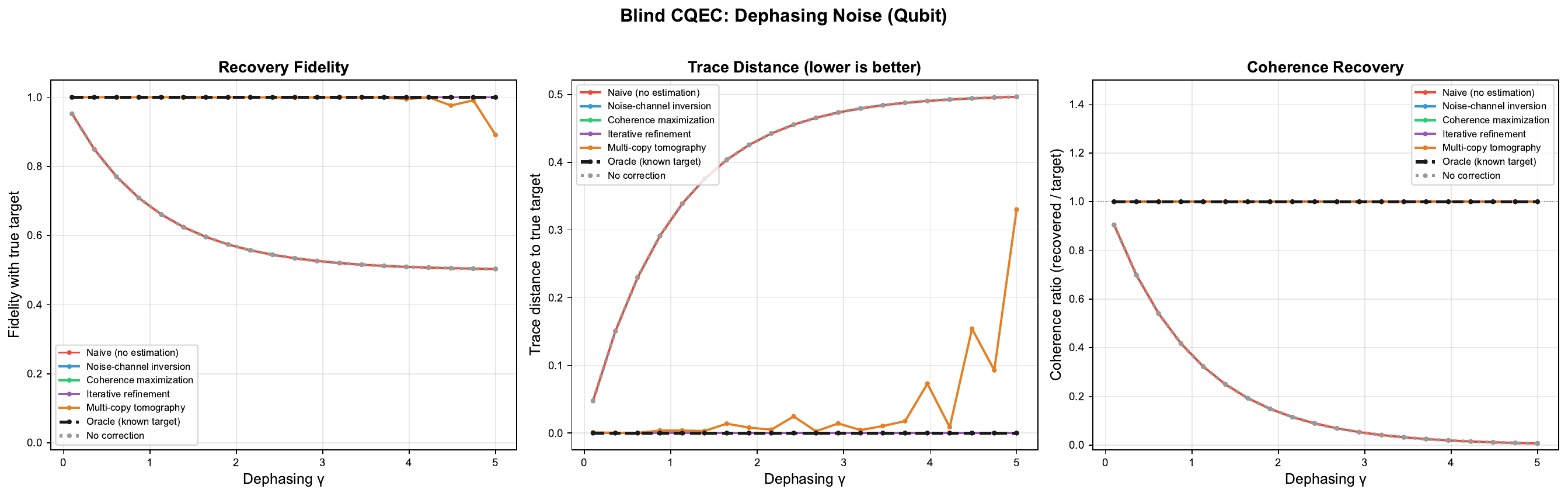}
\caption{\textbf{Coherence maximization is exact under pure dephasing.}
Recovery fidelity (left), trace distance (center), and coherence ratio (right) versus dephasing strength $\gamma$ for the five estimation strategies, the oracle (known target, dashed black), and the no-correction baseline.
Coherence maximization and iterative refinement track the oracle exactly across $\gamma \in [0.1, 5.0]$, while naive estimation never improves on the uncorrected state.}
\label{fig:noise_sweep}
\end{figure*}

Figure~\ref{fig:noise_sweep} shows the recovery fidelity as a function of noise strength for a maximally coherent qubit ($d=2$).

Under dephasing and depolarizing noise, coherence maximization, channel inversion, and iterative refinement all achieve $\Frec = 1.000$, matching the oracle exactly.
The key insight is that these channels preserve the phase structure $\phi_{ij} = \arg(\rho_{ij})$ while attenuating only the magnitudes---exactly the information that coherence maximization restores.

\begin{figure}[!tb]
\centering
\includegraphics[width=\columnwidth]{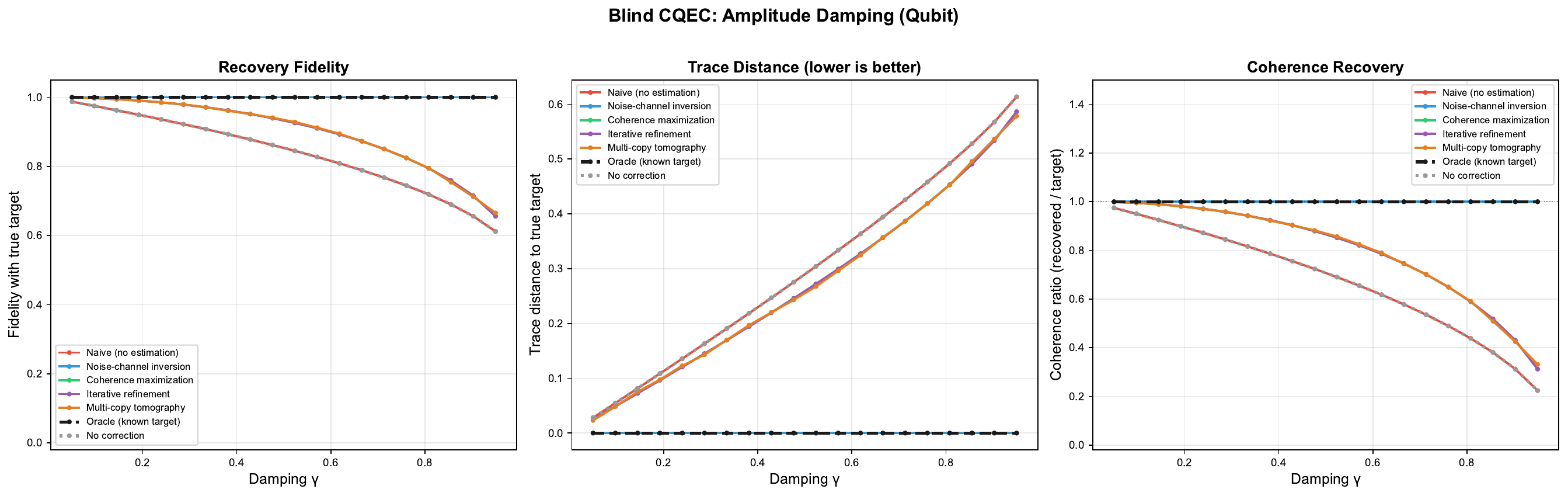}
\caption{\textbf{Coherence maximization fails under amplitude damping; only channel inversion is exact.}
Same axes as Fig.~\ref{fig:noise_sweep}, with damping rate $\gamma_\mathrm{AD} \in [0.05, 0.95]$.
Channel inversion (blue) tracks the oracle; coherence maximization (green) saturates due to population redistribution that cannot be inferred from off-diagonal data alone.}
\label{fig:ampdamp}
\end{figure}

Amplitude damping (Fig.~\ref{fig:ampdamp}) breaks this equivalence: only channel inversion achieves $\Frec = 0.9999$ (matching Oracle), while coherence maximization saturates at $\Frec = 0.926$ at $\gamma_\mathrm{AD} = 0.5$.
The failure is due to population redistribution ($\rho_{11} \to (1-\gamma)\rho_{11}$), which cannot be inferred from the off-diagonal structure.

\subsection{Four quantum algorithms}
\label{sec:algorithms_results}

Table~\ref{tab:main_results} presents the main results across all 4 algorithms under combined noise, with 5 estimation strategies plus no-correction baseline and oracle (all computed with the unified noise module and mode-restricted recovery; $n_\mathrm{copies} = 100$).

\begin{table*}[t]
\caption{\textbf{Exactly calibrated channel inversion matches the oracle on every algorithm; coherence maximization follows within $0.7$--$21\%$; the Regev ceiling of $0.77$ is set by the $10^{-10}$ mode threshold, not by estimation error.}
Recovery fidelity $\Frec = F(\rhor, \rhot)$ under combined noise ($\gamma = 1.0$, $p = 0.15$, $\gamma_\mathrm{AD} = 0.1$).
Bold values indicate the best blind strategy per algorithm.}
\label{tab:main_results}

\setlength{\tabcolsep}{3pt}
\begin{tabular}{lccccccc}
\hline
Algorithm ($d$) & No corr. & Naive & Ch.\ inv. & Coh.\ max & Iter. & Multi-c. & Oracle \\
\hline
qDRIFT ($d=8$)            & 0.195 & 0.195 & \textbf{1.000} & 0.986 & 0.884 & 0.536 & 1.000 \\
QKAN ($d=4$)              & 0.380 & 0.380 & \textbf{1.000} & 0.993 & 0.959 & 0.964 & 1.000 \\
Control-free QPE ($d=16$) & 0.268 & 0.268 & \textbf{1.000} & 0.992 & 0.889 & 0.335 & 1.000 \\
Regev factoring ($d=64$)  & 0.369 & 0.369 & \textbf{0.773} & 0.610 & 0.610 & 0.043 & 0.773 \\
\hline
\end{tabular}

\end{table*}

Three features of Table~\ref{tab:main_results} carry the physics.
First, channel inversion equals the oracle to every reported digit on every algorithm---the consistency signature of an exact forward/inverse pair (Sec.~\ref{sec:haar_sweep}).
Second, coherence maximization, requiring no noise model, lands within $0.7$--$1.4\%$ of the oracle at $d \leq 16$ and within $21\%$ at $d = 64$.
Third, the Regev ceiling of $0.773$ (shared by the oracle) is a property of the recovery interface, not of the estimators: under the combined noise the wide-gap coherences of the Regev state fall below the mode threshold $\tau_\mathrm{mode} = 10^{-10}$ and are pruned by $\mathcal{M}_{\rhon}$, so no estimator---including the oracle---can exceed it at this threshold (its dependence on $\tau_\mathrm{mode}$ is tabulated in Sec.~\ref{sec:computational}).

\begin{figure}[!tb]
\centering
\includegraphics[width=\columnwidth]{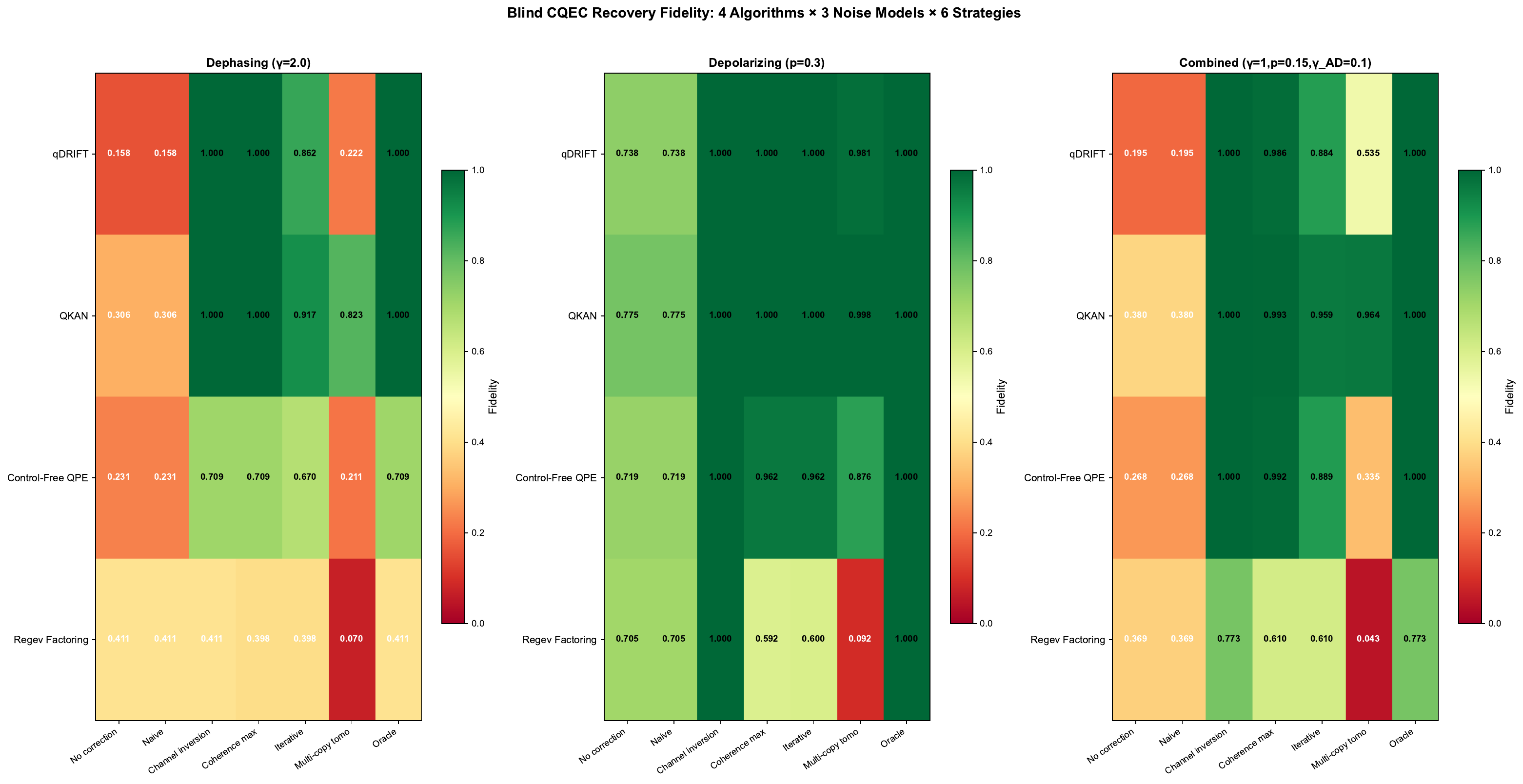}
\caption{\textbf{Channel inversion tracks the oracle in every cell; the visible dimension dependence is the recovery ceiling, not strategy ranking.}
Heatmap of recovery fidelity across 4~algorithms $\times$ 3~noise models $\times$ 5~estimation strategies (plus no-correction baseline and oracle).
Dark green indicates high fidelity; red indicates low fidelity.}
\label{fig:heatmap}
\end{figure}

To decompose the combined-noise results, Tables~\ref{tab:dephasing_only}--\ref{tab:ampdamp_only} present per-channel results for representative pure targets (one fixed Haar-random state per dimension), computed with the same unified noise module (Sec.~\ref{sec:computational}).
For the actual algorithm states the same run gives a sharper boundary case worth reporting: under dephasing-only noise ($\gamma = 2.0$) every off-diagonal element of the Regev state falls below the mode threshold $\tau_\mathrm{mode} = 10^{-10}$ (the 380 nonzero elements that remain range from $10^{-69}$ to $2\times10^{-12}$), so the interface prunes all of them and every strategy---oracle included---returns the no-correction fidelity ($0.41$).
This is a limit of the thresholded interface, not of Theorem~1, which places no lower bound on the magnitude of a recoverable coherence; lowering $\tau_\mathrm{mode}$ lifts this ceiling (Sec.~\ref{sec:computational}).

\begin{table}[!tb]
\caption{Recovery fidelity under dephasing only ($\gamma = 2.0$).
The Oracle column equals channel inversion at every entry: with exact parameters the inverted estimate is the true target, so the remaining deficit is the recovery-interface ceiling set by the mode threshold (coherences at large gaps fall below $\tau_\mathrm{mode}$ and are pruned by the interface; see Table~\ref{tab:tau_sensitivity}).}
\label{tab:dephasing_only}

\begin{tabular}{lcccc}
Strategy & $d{=}4$ & $d{=}8$ & $d{=}16$ & $d{=}64$ \\
\hline
No correction   & 0.414 & 0.279 & 0.117 & 0.036 \\
Coherence max   & \textbf{1.000} & \textbf{1.000} & \textbf{0.754} & 0.158 \\
Ch.\ inversion  & \textbf{1.000} & \textbf{1.000} & \textbf{0.754} & \textbf{0.207} \\
Oracle          & 1.000 & 1.000 & 0.754 & 0.207 \\
\end{tabular}

\end{table}

\begin{table}[!tb]
\caption{Recovery fidelity under depolarizing only ($p = 0.3$).
Depolarizing noise preserves the coherence-mode structure, so exact inversion is perfect at every dimension.}
\label{tab:depol_only}

\begin{tabular}{lcccc}
Strategy & $d{=}4$ & $d{=}8$ & $d{=}16$ & $d{=}64$ \\
\hline
No correction   & 0.775 & 0.738 & 0.719 & 0.705 \\
Coherence max   & 0.944 & 0.975 & 0.971 & 0.968 \\
Ch.\ inversion  & \textbf{1.000} & \textbf{1.000} & \textbf{1.000} & \textbf{1.000} \\
Oracle          & 1.000 & 1.000 & 1.000 & 1.000 \\
\end{tabular}

\end{table}

\begin{table}[!tb]
\caption{Recovery fidelity under amplitude damping only ($\gamma_\mathrm{AD} = 0.3$).}
\label{tab:ampdamp_only}

\begin{tabular}{lcccc}
Strategy & $d{=}4$ & $d{=}8$ & $d{=}16$ & $d{=}64$ \\
\hline
No correction   & 0.814 & 0.400 & 0.173 & 0.040 \\
Coherence max   & 0.965 & 0.878 & 0.838 & 0.689 \\
Ch.\ inversion  & \textbf{1.000} & \textbf{1.000} & \textbf{1.000} & \textbf{0.793} \\
Oracle          & 1.000 & 1.000 & 1.000 & 0.793 \\
\end{tabular}

\end{table}

The per-channel decomposition supports three observations.
(i)~\emph{The recovery-interface ceiling, not estimation, dominates at high dimension for gap-dependent channels}: under dephasing-only and amplitude-damping-only noise, even the oracle degrades with $d$ (to $0.21$ under dephasing and $0.79$ under amplitude damping at $d = 64$) because coherences at large mode gaps fall below the mode threshold and are pruned by the interface for every estimator (a threshold convention, Sec.~\ref{sec:bottleneck}, not a consequence of Theorem~1).
(ii)~\emph{Under depolarizing noise the mode structure survives} and exact inversion is perfect at every dimension, with coherence maximization within $0.06$ of it ($\geq 0.94$).
(iii)~\emph{Coherence maximization matches exact channel inversion wherever the phase structure survives} (dephasing at $d \leq 16$, to every reported digit) and falls behind where population information degrades (dephasing at $d = 64$: $0.158$ vs.\ $0.207$; amplitude damping at all $d$, e.g.\ $0.838$ vs.\ $1.000$ at $d = 16$).
With exactly calibrated parameters, channel inversion equals the oracle at every entry; its practical disadvantage appears only under parameter misspecification (Sec.~\ref{sec:sensitivity}).

\subsection{Estimation--recovery correlation}
\label{sec:correlation}

\begin{figure}[!tb]
\centering
\includegraphics[width=\columnwidth]{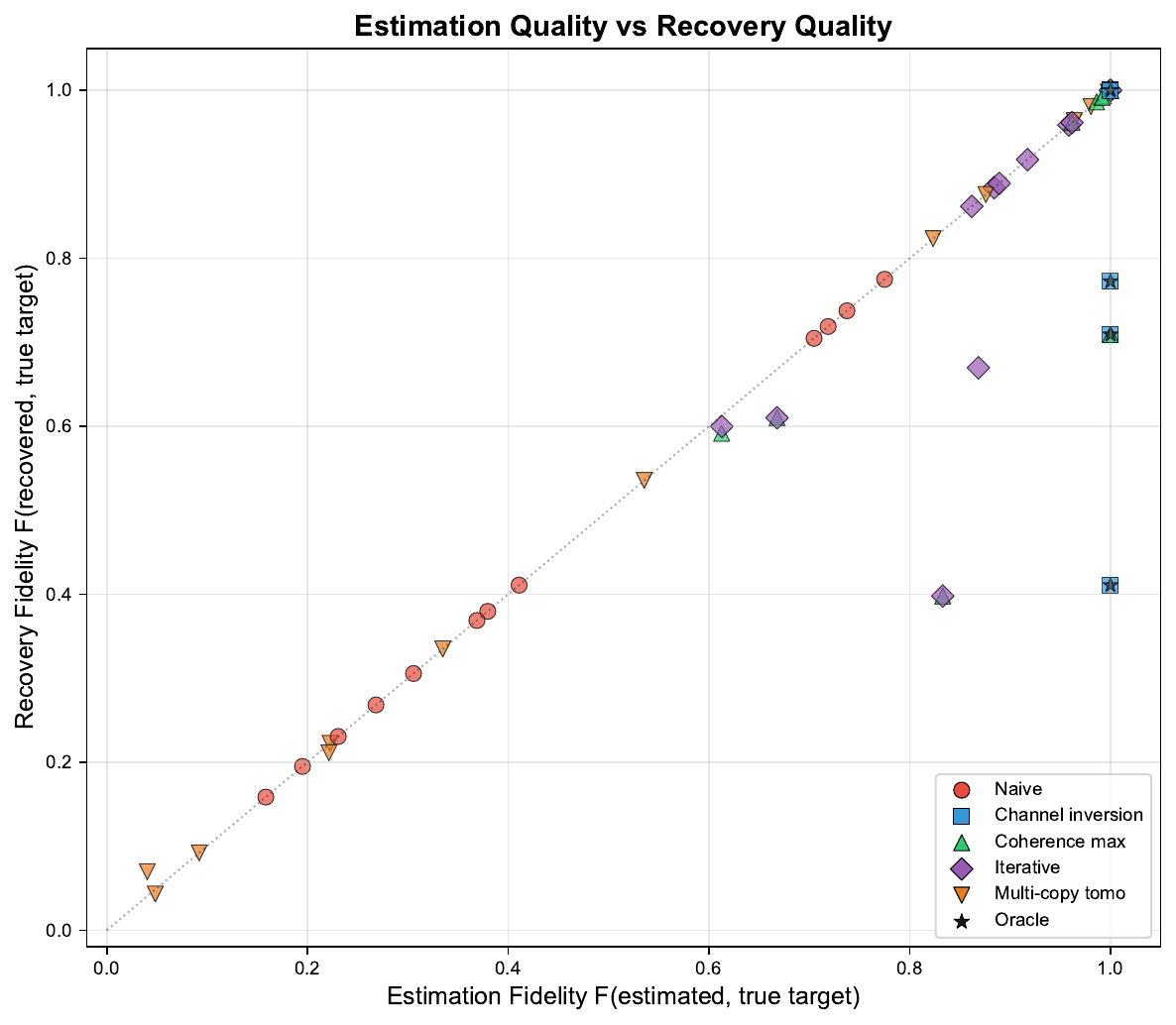}
\caption{\textbf{Recovery fidelity tracks estimation fidelity up to the mode-restriction penalty $\Delta_\mathrm{mode}$, exactly as Theorem~2 [Eq.~\eqref{eq:explicit_bound}] requires: on-diagonal where all modes survive, and below the diagonal where mode annihilation caps recovery.}
Estimation fidelity vs.\ recovery fidelity for all 84 algorithm--noise--condition combinations (Pearson $r$: $1.000$ at $d = 4$ and $d = 8$, $0.942$ at $d = 16$, $0.774$ at $d = 64$; pooled $0.920$).
The largest deviation ($\Fest = 1.00$ vs.\ $\Frec = 0.41$; Regev under dephasing-only) is a perfect estimate of a largely unrecoverable state.}
\label{fig:scatter}
\end{figure}

Figure~\ref{fig:scatter} plots estimation fidelity $\Fest = F(\rhoe, \rhot)$ against recovery fidelity $\Frec$ across all 84 data points (4 algorithms $\times$ 3 noise models $\times$ 7 conditions: 5 strategies plus baseline and oracle).

We are explicit about what this figure does and does not show.
Because the recovery interface is $\rhor = \Pi(\mathcal{M}_{\rhon}(\rhoe))$ [Eq.~\eqref{eq:recovery_structure}], the relationship between the two fidelities is fixed by construction: $\Frec = \Fest$ whenever the mode-restriction penalty $\Delta_\mathrm{mode}$ vanishes, and $\Frec$ falls below $\Fest$ by (at most) $\Delta_\mathrm{mode}$ otherwise.
The data conform to this structure quantitatively.
At low dimension, where all modes survive, the points sit on the diagonal ($r = 1.000$ at $d = 4$ and $d = 8$); as dimension grows and gap-dependent dephasing annihilates wide-gap coherences, the correlation degrades exactly as the bound predicts ($r = 0.942$ at $d = 16$, $0.774$ at $d = 64$), with the extreme case---a \emph{perfect} estimate ($\Fest = 1.00$) recovering only $\Frec = 0.41$ for the Regev state under dephasing-only noise---realizing the maximal mode-restriction penalty in our data set.
The figure is therefore a quantitative verification of Theorem~2 in both of its regimes [Eq.~\eqref{eq:fidelity_penalty} where all modes survive, Eq.~\eqref{eq:explicit_bound} where they do not], not an independent empirical discovery about catalytic dynamics; within this framework, performance differences between strategies originate in the estimation step, and the recovery ceiling is set by mode survival.
Whether a physical (circuit-level) catalytic implementation preserves this structure is an open question that cannot be answered at the density-matrix level; the pooled regression ($\Frec = 0.91\,\Fest + 0.02$, $R^2 = 0.85$; the 84 points are not i.i.d.\ and the statistics are descriptive) summarizes this benchmark ensemble only.

\subsection{Copy scaling}
\label{sec:copy_scaling}

\begin{figure*}[t]
\centering
\includegraphics[width=0.95\textwidth]{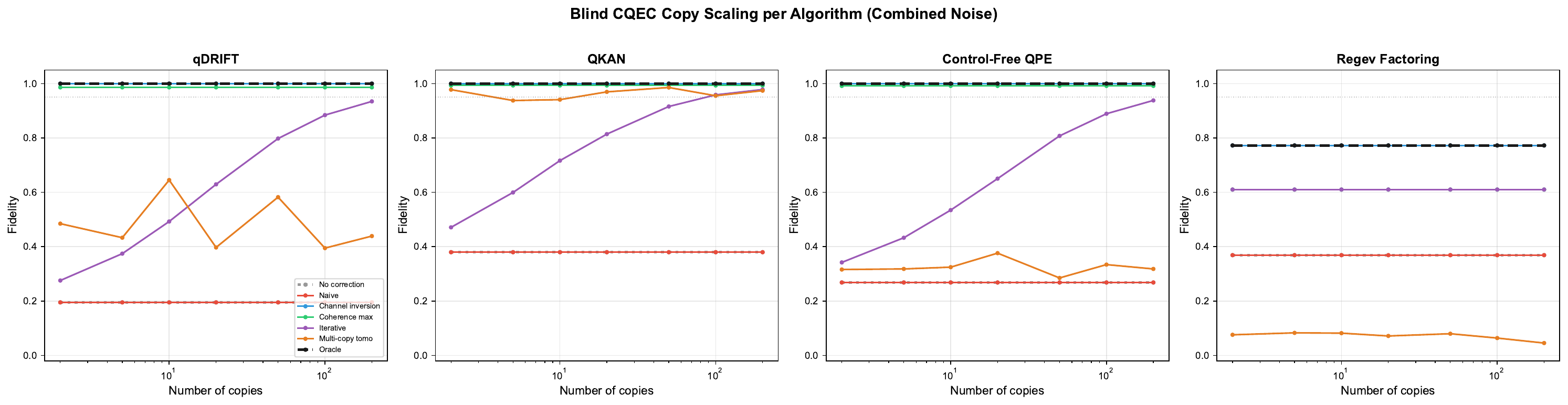}
\caption{\textbf{Single-representation strategies (channel inversion, coherence maximization, oracle) are independent of the copy count; only the iterative and multi-copy strategies move with $n$; at $d = 64$ no strategy---including the oracle---reaches $\Frec \geq 0.95$, because the mode-restricted ceiling ($0.77$) lies below it.}
Fidelity vs.\ copy count $n_\mathrm{copies}$ for the four quantum algorithms under combined noise.
Each panel shows the five blind strategies plus the no-correction baseline.}
\label{fig:copies}
\end{figure*}

Figure~\ref{fig:copies} shows $\Frec$ versus copy count $n_\mathrm{copies}$ for each algorithm under combined noise.

\begin{table}[b]
\caption{\textbf{Minimum state preparations for $\Frec \geq 0.95$ per algorithm (combined noise); ``$>200$'' = threshold not reached.}
``Preparations'' counts noisy-state inputs under the fluctuation-averaging model of Sec.~\ref{sec:copy_scaling}, conditional on oracle density-matrix access (Sec.~\ref{sec:qem_compare}); for strategies that consume one state representation (channel inversion, coherence maximization), averaging is applied as preprocessing.
Channel inversion, coherence maximization, and the oracle are independent of $n$ and meet the threshold at the smallest value of the grid ($n = 2$) wherever they meet it at all; Regev ($d = 64$) never reaches the threshold because the mode-restricted ceiling ($0.77$) lies below it.}
\label{tab:min_copies}

\begin{tabular}{lcccc}
Strategy & qDRIFT & QKAN & CF-QPE & Regev \\
\hline
Channel inversion & 2 & 2 & 2 & $>200$ \\
Coherence max     & 2 & 2 & 2 & $>200$ \\
Iterative         & $>200$ & 100 & $>200$ & $>200$ \\
Multi-copy avg.   & $>200$ & 2 & $>200$ & $>200$ \\
Oracle            & 2 & 2 & 2 & $>200$ \\
\end{tabular}

\end{table}

Under the recovery interface of Eq.~\eqref{eq:recovery_structure} the copy count enters only through the estimator.
Channel inversion, coherence maximization, and the oracle consume a single state representation and are therefore exactly constant in $n$ (Fig.~\ref{fig:copies}); iterative refinement, whose damped fixed point is reached deterministically, is likewise $n$-independent in the qubit and qutrit sweeps and gains with $n$ only on the algorithm states, where it approaches coherence maximization from below ($0.28 \to 0.93$ for qDRIFT between $n = 2$ and $200$).
Multi-copy averaging is the one strategy whose fidelity is a genuine function of $n$, and fitting $F(n) = 1 - A\,n^{-\alpha}$ to it by nonlinear least squares gives $\alpha = 0.87$ for the qubit under dephasing at $\gamma = 3$, $\alpha = 0.23$ and $0.13$ for the qutrit at $\gamma = 1$ and $3$, and $\alpha \approx 0$ under amplitude damping at $\gamma_\mathrm{AD} = 0.7$, where averaging cannot undo the population transfer.
Two methodological caveats limit the physical interpretation of these exponents.
First, the multi-copy ensemble is generated by adding fixed-amplitude Gaussian matrix perturbations ($\sigma = 0.01$ per element) to the noisy state, a proxy for state-preparation fluctuations that is \emph{not} derived from Born-rule measurement statistics; the absolute perturbation scale means the effective signal-to-noise ratio varies with dimension.
Second, the fit is to a single strategy on a seven-point grid, so the exponents are descriptive summaries of this simulation pipeline; establishing measurement-derived scaling laws requires shot-level simulation with explicitly allocated POVMs, which we identify as necessary future work (Sec.~\ref{sec:limitations}).
The operational content of this section is therefore negative: within the density-matrix framework, additional copies do not improve the two effective strategies, so any copy budget should be spent on acquiring one accurate representation of $\rhon$ rather than on averaging many.

\subsection{Dimension dependence}
\label{sec:dimension}

\begin{figure}[!tb]
\centering
\includegraphics[width=\columnwidth]{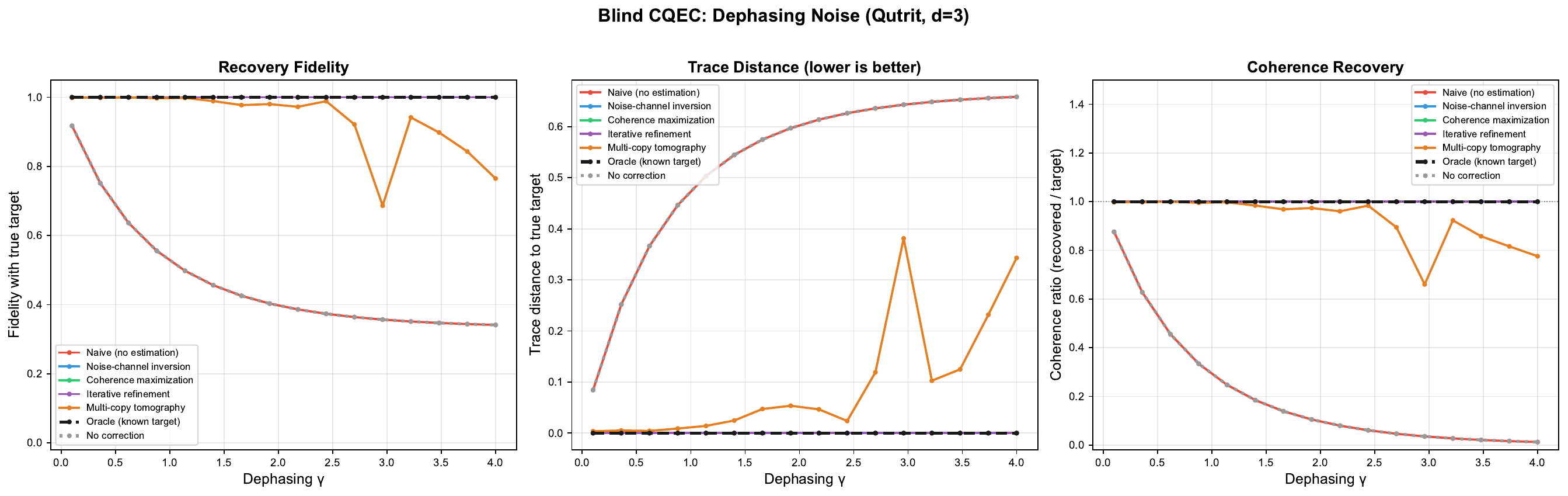}
\caption{\textbf{Coherence maximization tracks the oracle across the full dephasing range for a qutrit.}
Blind CQEC for a qutrit ($d = 3$) under dephasing, illustrating the intermediate behavior between qubit ($d = 2$) and the higher-dimensional algorithms.}
\label{fig:qutrit}
\end{figure}

The qutrit ($d = 3$) benchmarks confirm intermediate behavior between qubit ($d = 2$) and the higher-dimensional algorithms.
Figure~\ref{fig:qutrit} shows that coherence maximization tracks the oracle at $\Frec = 1.000$ over the whole dephasing range $\gamma \in [0.1, 4.0]$ for the qutrit, as for the qubit; at higher dimension the same strategy separates from the oracle under dephasing only once the mode threshold prunes wide-gap coherences (Table~\ref{tab:dephasing_only}: $0.754$ at $d = 16$ and $0.158$ at $d = 64$ for $\gamma = 2.0$).
The degradation is monotonic in $d$, consistent with the loosening of the physicality bound $\sqrt{p_i p_j} \to 1/d$ as $d \to \infty$.

\subsection{Dimension sweep with Haar-random states}
\label{sec:haar_sweep}

The algorithm-specific benchmarks of Secs.~\ref{sec:algorithms_results}--\ref{sec:dimension} used fixed target states determined by each quantum algorithm.
To assess strategy performance independent of target-state structure, we performed a systematic dimension sweep using Haar-random pure states (generated via \path{benchmark_blind_dimension_sweep.py}).
For each dimension $d \in \{2, 4, 8, 16, 32, 64\}$, 20~Haar-random pure states were drawn, subjected to combined noise ($\gamma = 1.0$, $p = 0.15$, $\gamma_\mathrm{AD} = 0.1$), and corrected using each estimation strategy.
(Dimensions $d \geq 128$ are excluded because the exact dephasing inversion factor $e^{\gamma(d-1)}$ exceeds double-precision range.)

\begin{figure}[!tb]
\centering
\includegraphics[width=\columnwidth]{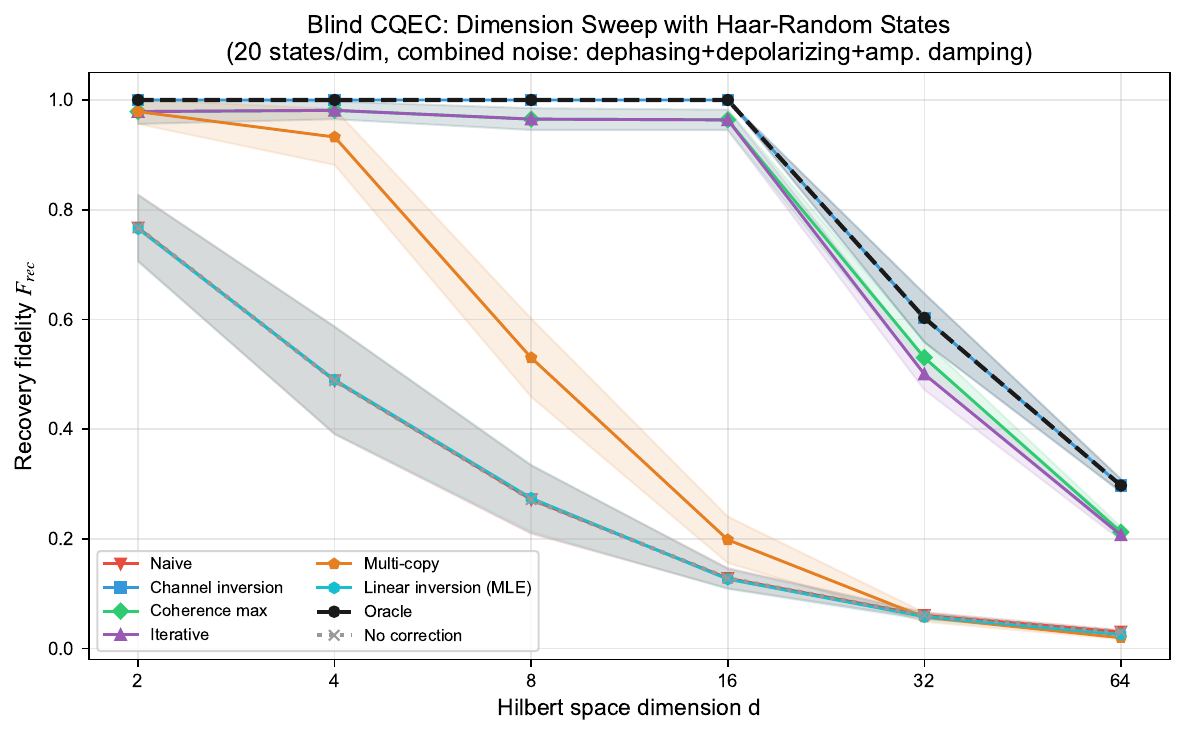}
\caption{\textbf{With exactly calibrated parameters, channel inversion coincides with the oracle at every dimension; both are capped by the recovery-interface ceiling, and coherence maximization falls progressively below them.}
Recovery fidelity versus Hilbert space dimension for Haar-random pure states under combined noise.
Each point is the mean over 20~Haar-random targets; shaded bands show $\pm 1\sigma$.
Channel inversion (= oracle) is exact ($\Frec = 1.000$) for $d \leq 16$ and drops to $0.603$ ($d = 32$) and $0.297$ ($d = 64$) once the mode threshold prunes wide-gap coherences, while coherence maximization runs below it throughout, from $0.979$ ($d = 2$) to $0.212$ ($d = 64$).}
\label{fig:dimension_sweep}
\end{figure}

Figure~\ref{fig:dimension_sweep} shows the mean recovery fidelity as a function of~$d$, and its structure is best read in two layers.
First, \emph{channel inversion with exact parameters coincides with the oracle to every reported digit at every dimension} ($1.000$ for $d = 2$--$16$, then $0.603$ and $0.297$ at $d = 32$ and $64$): this is the expected signature of a consistent forward/inverse pair, since exact inversion reproduces the true target, and the deficit that appears at $d \geq 32$ is the recovery-interface ceiling (wide-gap coherences pruned by the mode threshold under gap-dependent dephasing, cf.\ Table~\ref{tab:dephasing_only}), not estimation error.
Second, \emph{coherence maximization tracks the ceiling closely at low dimension and separates from it as $d$ grows} ($0.979 \to 0.212$), with the gap to channel inversion opening from $0.02$ at $d = 2$ to $0.09$ at $d = 64$; the gap is widest at $d = 32$ ($0.07$) and $d = 64$ where the ceiling itself has fallen.
Consequently, under exact calibration there is \emph{no} crossover: channel inversion is never worse.
The operationally meaningful crossover emerges only when calibration error is taken into account (Sec.~\ref{sec:sensitivity}): at $d = 64$, a $-10\%$ parameter error drops channel inversion to $\Frec = 0.18$, below coherence maximization's noise-model-free $0.21$.

\begin{figure}[!tb]
\centering
\includegraphics[width=\columnwidth]{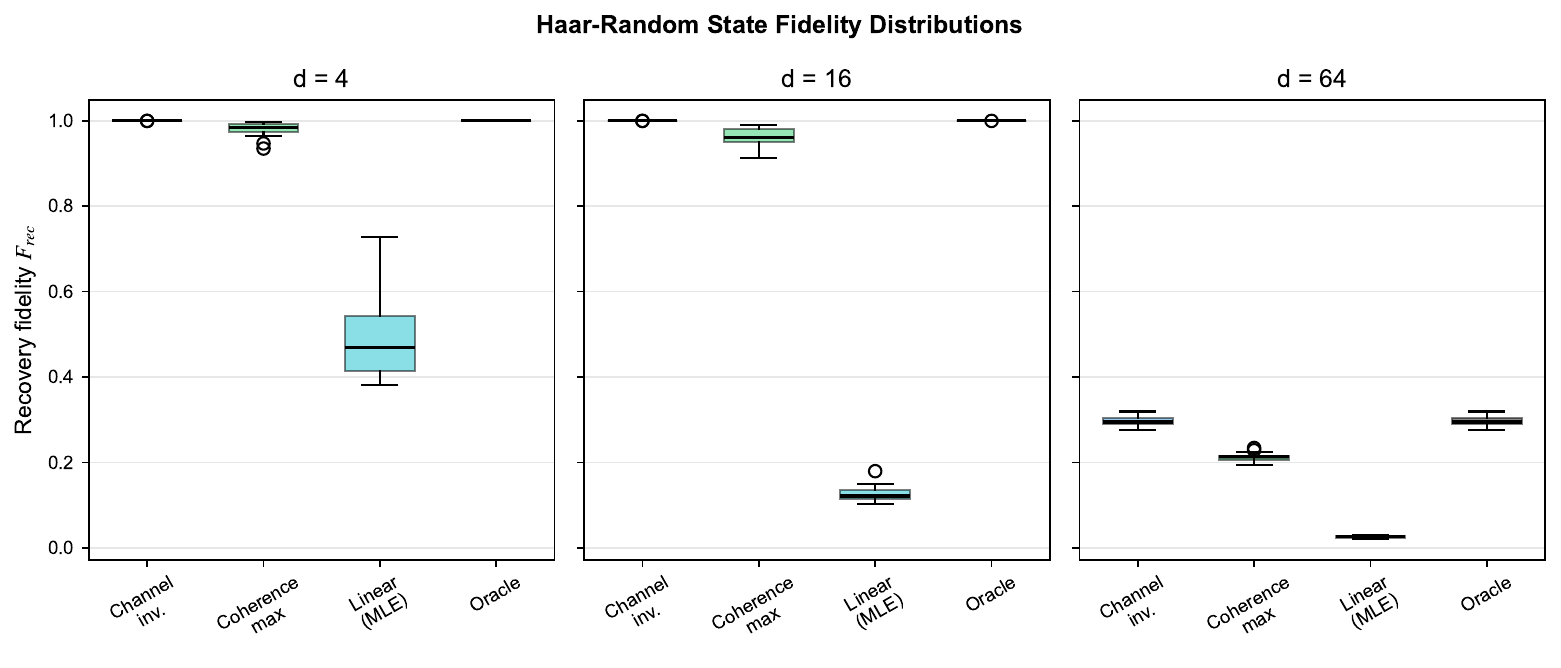}
\caption{\textbf{Concentration of measure tightens the fidelity distribution at large $d$ while the gap between strategies widens.}
Distribution of recovery fidelities for Haar-random states at selected dimensions.}
\label{fig:haar_distribution}
\end{figure}

The fidelity distributions across the 20 Haar-random targets concentrate with increasing~$d$ (standard deviations shrink from $0.10$ for the uncorrected state and $0.02$ for coherence maximization at $d = 4$ to $\leq 0.01$ for every strategy at $d = 64$), confirming that the reported means are representative (Fig.~\ref{fig:haar_distribution}).

\begin{figure}[!tb]
\centering
\includegraphics[width=\columnwidth]{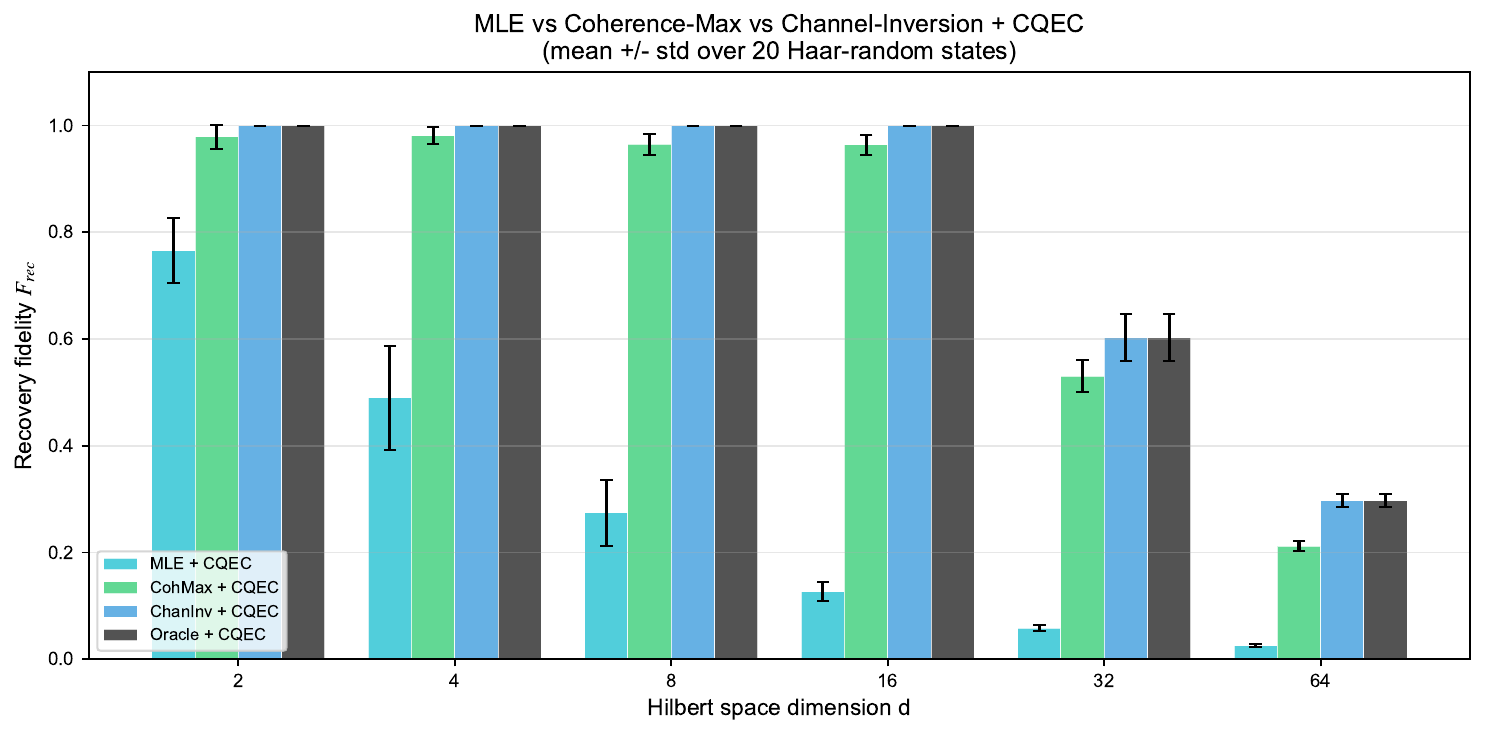}
\caption{\textbf{Linear-inversion tomography fails as a CQEC target estimator: it cannot substitute for decoherence-aware estimation.}
Comparison of blind CQEC strategies with standard linear inversion tomography (QST baseline) at $d = 8$.
Linear inversion (red dashed) performs at the no-correction level because it recovers $\EE(\rhot)$ rather than $\rhot$ itself.}
\label{fig:mle_comparison}
\end{figure}

A key finding is the comparison with standard linear inversion tomography.
Linear inversion---averaging $n$~noisy copies and projecting onto the positive semidefinite cone---recovers an estimate of $\EE(\rhot)$ rather than $\rhot$ itself.
When this estimate is used as the CQEC target, the recovery fidelity matches the no-correction baseline (Fig.~\ref{fig:mle_comparison}), because the systematic bias of the noise channel is not removed by averaging.
This validates the necessity of decoherence-aware estimation strategies (coherence maximization or channel inversion) for blind CQEC: generic quantum state tomography, which faithfully reconstructs the \emph{noisy} state, is insufficient because CQEC requires an estimate of the \emph{pre-noise} target.

\subsection{Noise-parameter sensitivity}
\label{sec:sensitivity}

Channel inversion (Sec.~\ref{sec:inversion}) requires knowledge of the noise parameters, which in practice must themselves be estimated.
To quantify the robustness of channel inversion to parameter misspecification, we performed a systematic sensitivity analysis (via \path{benchmark_blind_sensitivity.py}).

\begin{figure}[!tb]
\centering
\includegraphics[width=\columnwidth]{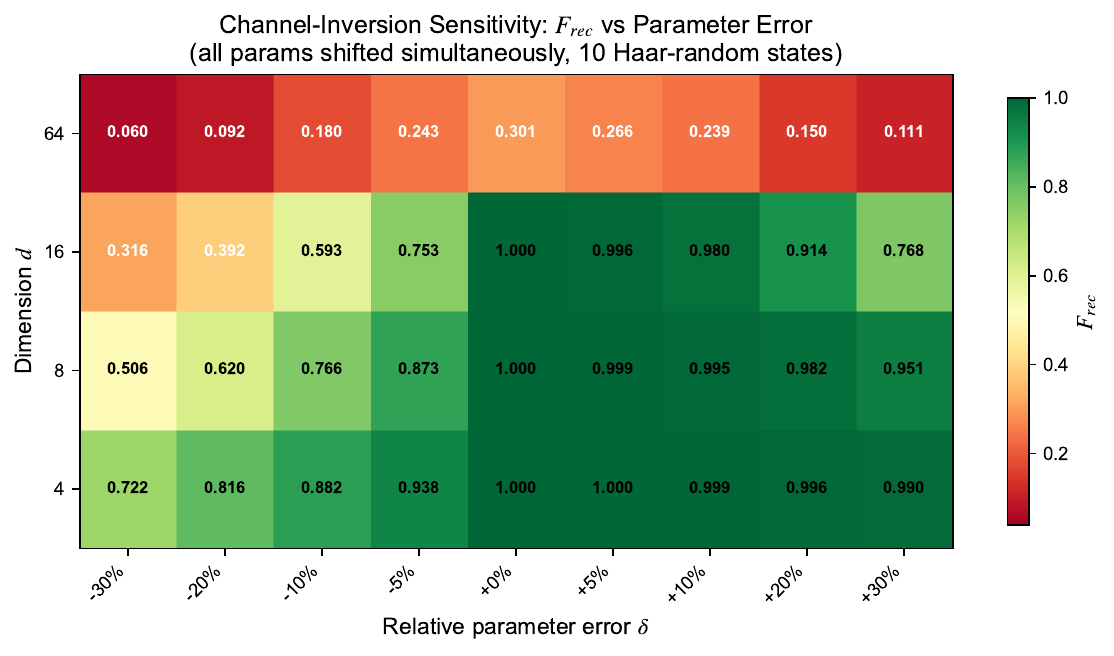}
\caption{\textbf{Channel inversion is sharply and asymmetrically sensitive to parameter misspecification: underestimating the noise is far more damaging than overestimating it, and the tolerance window narrows rapidly with dimension.}
Recovery fidelity $\Frec$ as a function of dimension~$d$ and relative parameter error~$\delta$ (all three noise parameters perturbed by the same relative fraction).
At $d = 4$, $\Frec \geq 0.72$ across $\delta \in [-30\%, +30\%]$; at $d = 64$, a $-5\%$ error already costs $19\%$ of the fidelity ($0.30 \to 0.24$), $-10\%$ costs $40\%$ ($\to 0.18$), and $+10\%$ costs $21\%$ ($\to 0.24$).}
\label{fig:sensitivity_heatmap}
\end{figure}

Figure~\ref{fig:sensitivity_heatmap} shows a heatmap of $\Frec$ versus $(d, \delta)$, where $\delta$ is the relative error applied uniformly to all three noise parameters.
At $d = 4$ the strategy is forgiving ($\Frec = 0.72$ at $\delta = -30\%$, $\geq 0.99$ for $\delta \geq 0$), but the tolerance window narrows quickly with dimension: at $d = 16$ a $-10\%$ error costs $0.41$ ($1.00 \to 0.59$), and at $d = 64$ the exact-parameter value $\Frec = 0.30$ falls to $0.24$ at $\delta = -5\%$, $0.18$ at $\delta = -10\%$, and $0.24$ at $\delta = +10\%$.
The degradation is strongly \emph{asymmetric}: underestimating the noise is far more damaging than overestimating it, because under-inversion leaves residual attenuation on exactly the wide-gap coherences that the mode-restriction step then removes permanently, whereas over-inversion errors are partially absorbed by the PSD projection.
The operational consequence is a direct comparison with the noise-model-free alternative: at $d = 64$, channel inversion beats coherence maximization ($\Frec = 0.21$) only when the calibration error lies within roughly $[-5\%, +10\%]$; outside this window the noise-model-free strategy is preferable despite its lower ceiling.

\begin{figure}[!tb]
\centering
\includegraphics[width=\columnwidth]{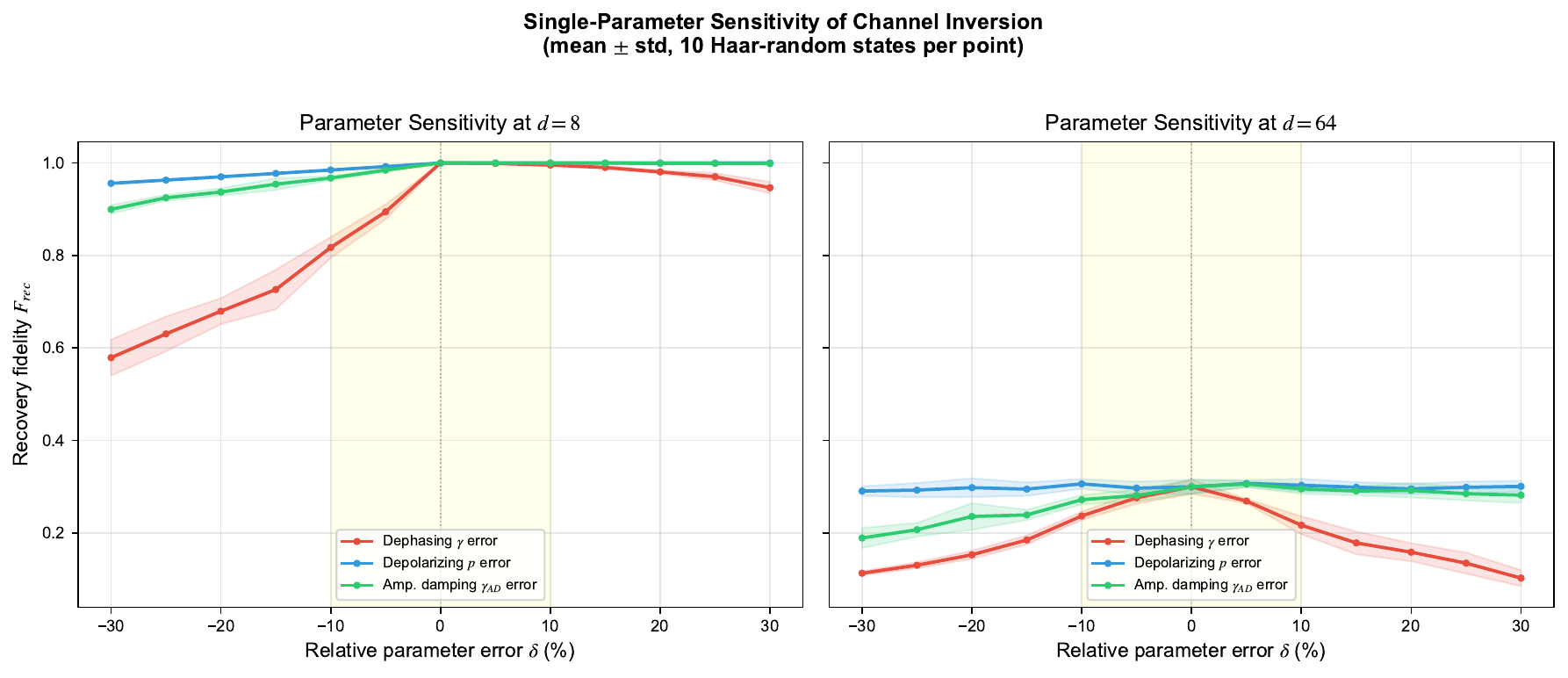}
\caption{\textbf{The dephasing rate $\gamma$ dominates the sensitivity budget; the depolarizing probability $p$ is nearly harmless to misspecify.}
Per-parameter sensitivity at $d = 8$ (solid) and $d = 64$ (dashed), varying one parameter at a time.
At $d = 64$: $-10\%$ error in $\gamma$ alone gives $\Frec = 0.24$ (vs.\ $0.30$ exact) and $-30\%$ gives $0.11$, while $\pm 30\%$ in $p$ leaves $\Frec$ essentially unchanged ($0.29$--$0.31$); $\gamma_\mathrm{AD}$ is intermediate.}
\label{fig:sensitivity_curves}
\end{figure}

Per-parameter analysis (Fig.~\ref{fig:sensitivity_curves}) shows that the error budget is dominated entirely by the dephasing rate~$\gamma$, which enters the inversion as $e^{+\gamma|\Delta_{ij}|}$ so that a relative error is amplified exponentially in the mode gap: at $d = 8$, a $-10\%$ error in $\gamma$ alone reduces $\Frec$ from $1.00$ to $0.82$ and $-30\%$ to $0.58$, whereas the depolarizing probability~$p$ can be misspecified by $\pm 30\%$ with almost no effect ($\geq 0.96$) and the damping rate $\gamma_\mathrm{AD}$ is intermediate ($0.90$ at $-30\%$); at $d = 64$ the same ordering holds ($\gamma$: $0.30 \to 0.11$ at $-30\%$; $p$: $0.29$--$0.31$ throughout; $\gamma_\mathrm{AD}$: $0.19$ at $-30\%$).
These results sharpen the calibration requirement considerably relative to a naive reading: at $d = 64$, randomized-benchmarking-grade characterization~\cite{Emerson2005,Knill2008,Magesan2011}---which estimates an average gate fidelity rather than a dephasing rate directly---must pin the \emph{dephasing rate specifically} to within a few percent (with a bias toward overestimation) for channel inversion to retain its advantage; at $d \leq 8$ a $\pm 10\%$ tolerance keeps $\Frec \geq 0.77$ and a $+10\%$ bias keeps it above $0.99$.

\subsection{Mixed-state targets}
\label{sec:mixed_state}

All preceding benchmarks used pure or near-pure target states.
To characterize blind CQEC on mixed-state targets, we generated Werner-like states $\rho(v) = v|\psi\rangle\!\langle\psi| + (1{-}v)I/d$ at $d = 8$ with purity parameter $v \in [0.3, 1.0]$ and 10~Haar-random $|\psi\rangle$ per~$v$.

\begin{figure}[!tb]
\centering
\includegraphics[width=\columnwidth]{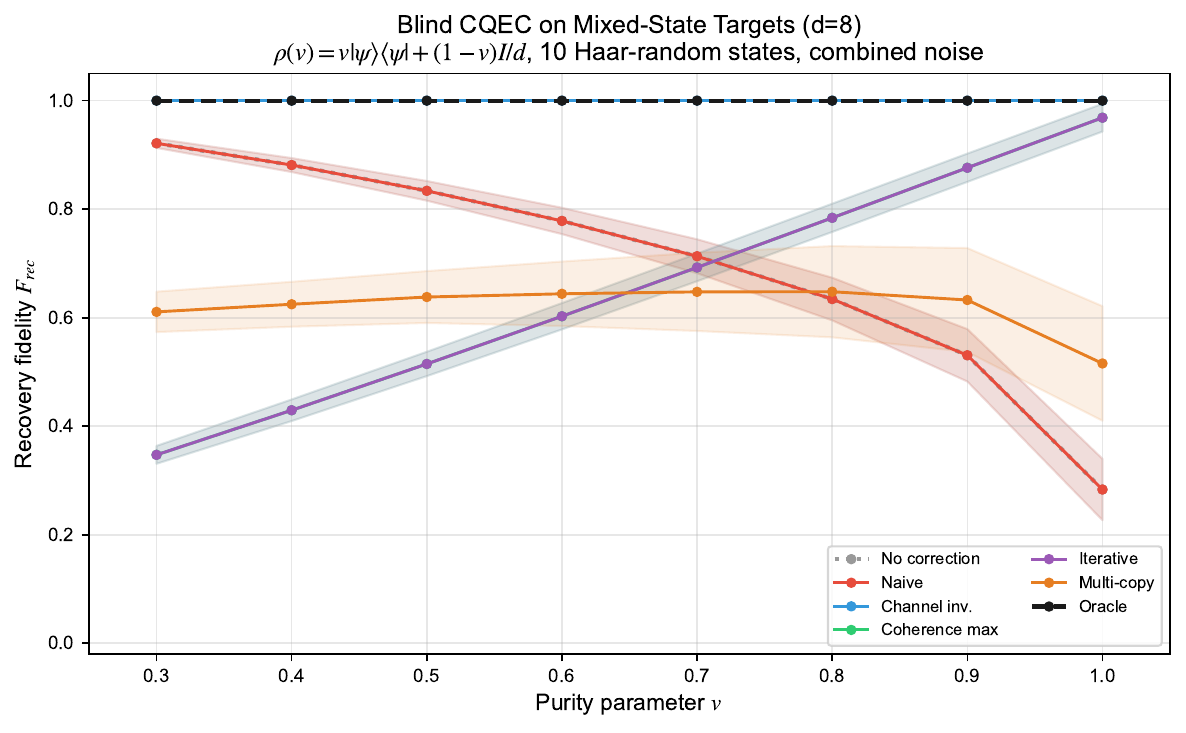}
\caption{\textbf{Coherence maximization becomes actively harmful for strongly mixed targets, while channel inversion is purity-robust.}
Recovery fidelity versus target-state purity $v$ for Werner-like states at $d = 8$ under combined noise.
For $v \lesssim 0.7$, coherence maximization falls below the no-correction baseline: the maximally-coherent ansatz injects coherence the mixed target does not possess.}
\label{fig:mixed_state}
\end{figure}

Figure~\ref{fig:mixed_state} reveals that coherence maximization degrades for mixed targets: the physicality bound $\sqrt{p_i p_j}$ overshoots the true coherences, biasing the estimate.
We report the operationally critical consequence explicitly: \emph{for $v \leq 0.7$, coherence maximization is worse than applying no correction at all} (at $v = 0.3$: corrected $\Frec = 0.35$ vs.\ no-correction $0.92$; at $v = 0.7$: $0.69$ vs.\ $0.71$), because the maximally-coherent ansatz actively injects coherence that the mixed target does not possess.
Any practical deployment of coherence maximization therefore requires a purity check before application; absent one, the strategy can degrade the state it is meant to protect.
Channel inversion, by contrast, is purity-robust: it is exact at every purity tested ($\Frec = 1.000$ at $v = 0.3$, $0.5$, $0.8$, and $1.0$), because at $d = 8$ under combined noise all modes survive the threshold and exact inversion reproduces the target whatever its purity.
This identifies mixed-state targets as a regime where noise-model knowledge is essential, even at low~$d$, and qualifies the low-dimensional recommendation of Sec.~\ref{sec:regimes}.

\subsection{Hybrid strategy}
\label{sec:hybrid}

The complementary strengths of coherence maximization (noise-model-free, effective at low~$d$) and channel inversion (noise-model-aware, effective at high~$d$) motivate a hybrid estimator:
\begin{equation}
\rhoe^{(\mathrm{hyb})}(w) = w\,\rhoe^{(\mathrm{inv})} + (1{-}w)\,\rhoe^{(\mathrm{CM})},
\label{eq:hybrid}
\end{equation}
where $w \in [0,1]$ interpolates between the two strategies.

\begin{figure}[!tb]
\centering
\includegraphics[width=\columnwidth]{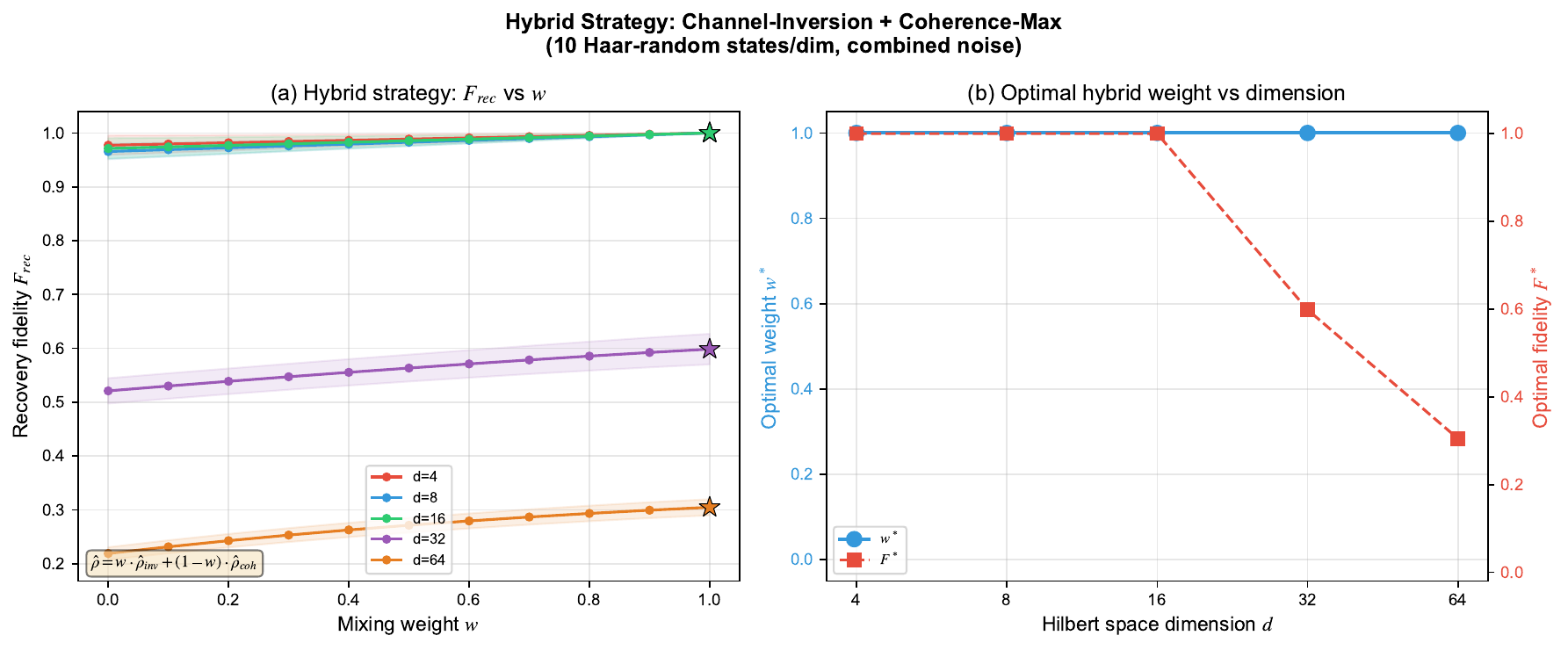}
\caption{\textbf{Under exact calibration the hybrid family is monotone in $w$ with $w^* = 1$ at every dimension: exact channel inversion is never improved by admixing coherence maximization.}
(a)~Recovery fidelity vs.\ mixing weight~$w$ at selected dimensions under combined noise ($w = 0$: coherence maximization; $w = 1$: channel inversion).
(b)~Optimal weight vs.\ dimension.
$w^*$ is selected by maximizing fidelity against the \emph{true} target (an oracle analysis; see text).}
\label{fig:hybrid}
\end{figure}

Figure~\ref{fig:hybrid} shows $\Frec$ versus~$w$ at each dimension under the unified noise implementation.
The result is unambiguous: $F(w)$ is monotonically increasing in $w$ at every dimension tested, so $w^* = 1$ (pure channel inversion) throughout, with no interior optimum.
This is the expected consequence of Sec.~\ref{sec:haar_sweep}: exact channel inversion reproduces the oracle, so no admixture of coherence maximization can improve on it.
The practical motivation for the hybrid family therefore rests entirely on \emph{parameter uncertainty}: when the calibration error is large enough that channel inversion underperforms coherence maximization (Sec.~\ref{sec:sensitivity}), an intermediate $w$ interpolates between the miscalibrated inversion and the noise-model-free fallback.
Two caveats apply: the $w^*$ reported here is selected by maximizing fidelity against the \emph{true} target (an oracle analysis characterizing the achievable envelope, not an operational tuning rule), and a blind selection rule for $w$ from calibration data alone remains future work.

\subsection{Comparison with standard quantum error mitigation}
\label{sec:qem_compare}

A natural question is how blind CQEC relates to established quantum error mitigation (QEM) methods for near-term devices: zero-noise extrapolation (ZNE)~\cite{Temme2017,Li2017,Kim2023utility}, probabilistic error cancellation (PEC)~\cite{Temme2017,Endo2018,vandenBerg2023pec}, and virtual distillation (VD)~\cite{Koczor2021,Huggins2021,OBrien2023purification}; see Ref.~\cite{Cai2023qem} for a review.
We preface the comparison with an important scope statement: \emph{the comparison below is not resource-matched}.
The standard QEM methods are designed to correct expectation values of observables from finite measurement data, with well-characterized sampling overheads; here we evaluate density-matrix-level idealizations of each method (ZNE as element-wise Richardson extrapolation of the density matrix, PEC as exact channel inversion without quasiprobability sampling, VD as $\rho^2/\tr\rho^2$ without a measurement protocol) on the same Haar-random states under combined noise (15 states per dimension).
The comparison therefore probes only one axis---what state-level information each idealized method can reconstruct---and says nothing about shot budgets, calibration cost, or bias--variance trade-offs, which would require a shot-level study.

\begin{figure}[!tb]
\centering
\includegraphics[width=\columnwidth]{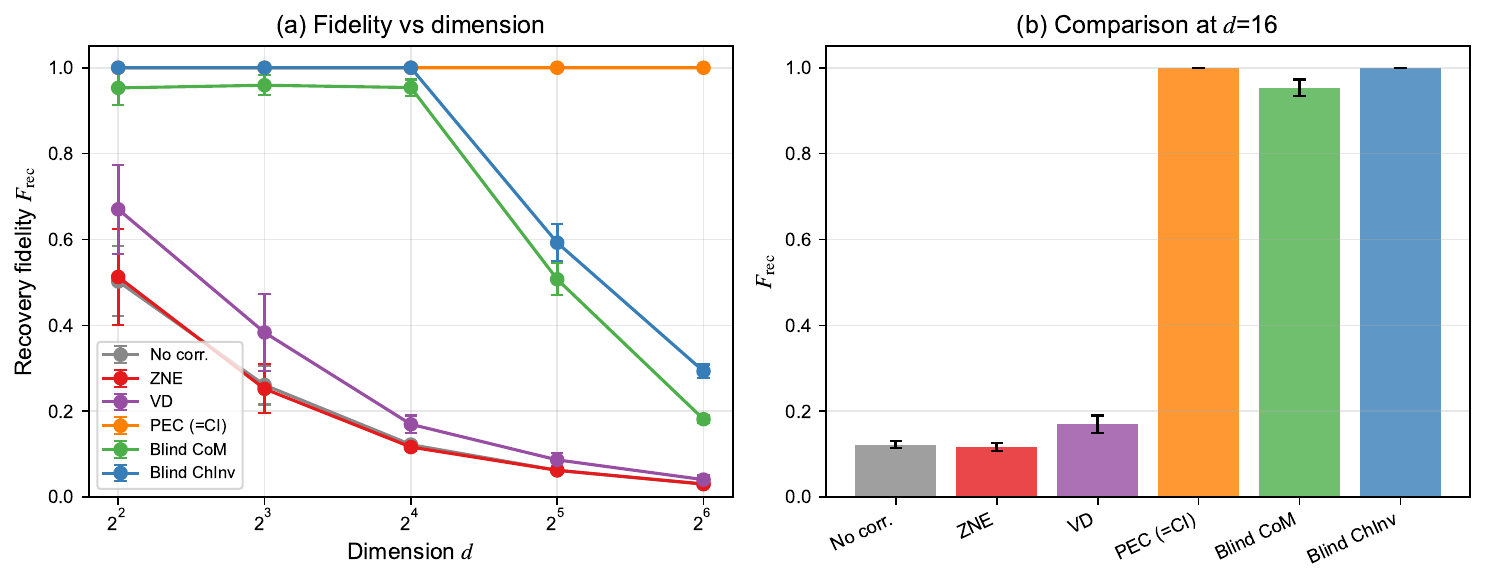}
\caption{\textbf{Idealized density-matrix-level comparison: exact-inversion methods (PEC, blind ChInv) reconstruct the state wherever the mode structure survives; extrapolation- and purification-based idealizations (ZNE, VD) do not; blind CoM is the only noise-model-free method that does.}
(a)~Recovery fidelity $\Frec$ vs.\ dimension for no correction, ZNE, VD, PEC, blind coherence maximization (CoM), and blind channel inversion (ChInv), all as density-matrix-level idealizations (see scope statement in text; the comparison is not resource-matched).
(b)~Bar chart at $d = 16$.}
\label{fig:qem_compare}
\end{figure}

Figure~\ref{fig:qem_compare} supports three observations within this idealized setting:
\begin{itemize}
\item \emph{Idealized PEC and blind ChInv coincide at $\Frec = 1.000$ for $d \leq 16$}, since with exactly known noise parameters both amount to exact channel inversion.
At $d = 32$ and $64$ they separate ($0.593$ and $0.293$ for blind ChInv against $1.000$ for idealized PEC): the PEC idealization inverts the channel on the full matrix, whereas blind ChInv passes its estimate through the recovery interface of Eq.~\eqref{eq:recovery_structure}, whose mode restriction prunes the wide-gap coherences that the noise has driven below $\tau_\mathrm{mode}$.
The gap is therefore the mode-survival penalty of Theorem~2, not an estimation error, and it is specific to the thresholded interface (Sec.~\ref{sec:computational}); the further distinction between the two methods lies in the operational layer not simulated here (PEC reconstructs observables via quasiprobability sampling with overhead exponential in circuit volume~\cite{Temme2017,Endo2018,Takagi2022limits,Quek2024tighter}, whereas a physical CQEC implementation would act on the state via a catalytic map whose copy cost is itself exponential in system size in the analysis of Ref.~\cite{Wakaura2026unified}; neither cost is simulated here).
\item \emph{The ZNE and VD idealizations scale poorly with dimension} ($\Frec \approx 0.51$ and $0.67$ at $d = 4$, collapsing to $0.03$--$0.04$ at $d = 64$), reflecting that extrapolation and purification do not reconstruct the pre-noise state when noise strongly mixes populations.
This is a statement about the state-reconstruction axis only---not a criticism of these methods on their design objective (expectation-value correction).
\item \emph{Blind CoM is the only method in this comparison that improves the state without noise-model knowledge}, reaching $\Frec > 0.95$ for $d \leq 16$; at $d = 64$ it degrades to $0.18$ (blind ChInv: $0.29$; idealized PEC: $1.000$), so its advantage is confined to the low-dimensional regime.
\end{itemize}

\begin{table}[!tb]
\caption{Classical post-processing cost and input requirements at $d = 8$ for the density-matrix-level idealizations (runtime is mean per recovery call in our simulation).
``State preparations'' counts the noisy-state inputs consumed per recovery \emph{given oracle density-matrix access}; it excludes the measurement cost of acquiring the density matrix, which for tomographic acquisition scales as $\tilde\Theta(d^2/\epsilon^2)$ copies with collective measurements (Sec.~\ref{sec:sample_complexity}) and dominates all entries in this table in any operational deployment.}
\label{tab:overhead}

\begin{tabular}{lccc}
Method & Runtime (ms) & State preps. & Noise info \\
\hline
Blind CoM      & 0.13 & 1 & None \\
Blind ChInv    & 0.08 & 1 & Full model \\
PEC (idealized) & 0.06 & 1 & Full model \\
VD (idealized)  & 0.02 & 2 & None \\
ZNE (idealized) & 1.36 & 3 & None \\
\end{tabular}

\end{table}

Table~\ref{tab:overhead} summarizes the classical costs of the idealized pipelines.
Within this scope, blind CoM occupies the niche of noise-model-free state-level improvement at low dimension; a resource-matched operational comparison (equal circuits, observables, total shots, and calibration budgets for all methods) is required before any deployment-level claim can be made, and we identify it as future work.

\subsection{Sample-complexity context}
\label{sec:sample_complexity}

The relevant information-theoretic reference point is the sample complexity of quantum state tomography: reconstructing an unknown $d$-dimensional state to trace-norm accuracy $\epsilon$ requires $\tilde\Theta(d^2/\epsilon^2)$ copies with collective measurements~\cite{Haah2017,ODonnellWright2016}, and $\Omega(d^3/\epsilon^2)$ up to logarithmic factors when the copies are measured independently~\cite{Haah2017}.
This bound frames the resource accounting of Sec.~\ref{sec:qem_compare}: because our simulations grant the estimators oracle access to the exact noisy density matrix, the copy counts reported in Table~\ref{tab:min_copies} quantify only the \emph{state-preparation fluctuation-averaging} requirement within our perturbation model, not the measurement cost of acquiring $\rhon$ in the first place.
A full operational analysis---allocating explicit POVMs and shot budgets to each estimator and comparing the resulting end-to-end sample complexity against the tomographic baseline---is a necessary next step that the density-matrix-level framework of this paper cannot provide.

\begin{figure}[!tb]
\centering
\includegraphics[width=\columnwidth]{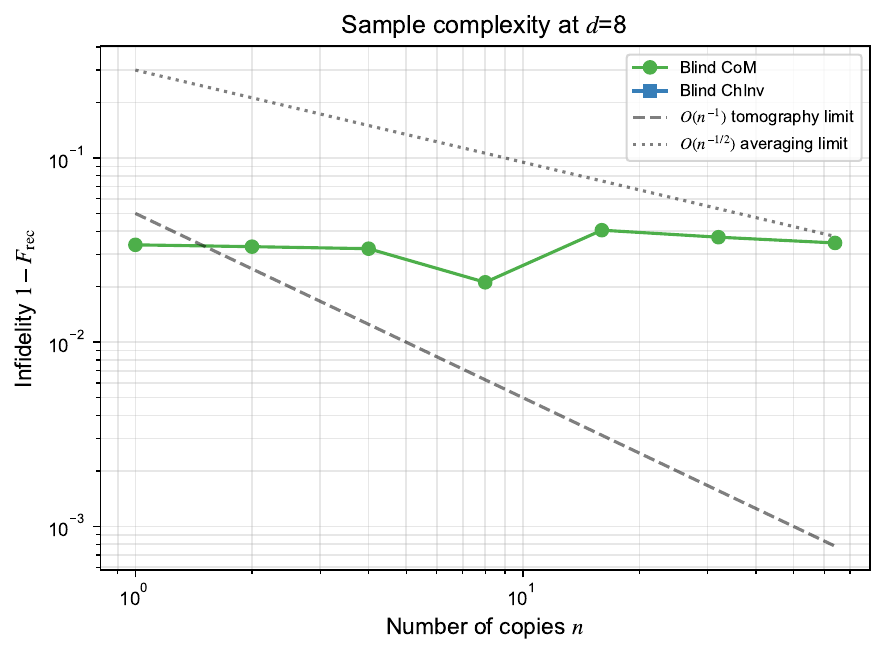}
\caption{\textbf{Empirical infidelity--copy curves at $d = 8$ are flat for both single-representation strategies; the $O(n^{-1})$ and $O(n^{-1/2})$ reference slopes are visual guides with illustrative prefactors.}
Because the copy ensemble derives from a Gaussian perturbation model rather than measurement statistics, these curves characterize the simulation pipeline; they do not establish saturation of information-theoretic bounds.}
\label{fig:sample_complexity}
\end{figure}

There is, moreover, a fundamental reason why oracle density-matrix access cannot be replaced by a single physical copy for any \emph{deterministic, target-independent} recovery.
Let $\sigma_\pm = (I \pm X)/2$ and let dephasing produce $\rho_\pm = (I \pm \eta X)/2$ with $0 < \eta < 1$; both noisy states retain nonzero coherence, so mode inclusion holds.
Any fixed apparatus acting on one noisy copy together with a target-independent catalyst induces a CPTP map $R$, and trace-distance contractivity gives $D(R(\rho_+), R(\rho_-)) \leq D(\rho_+, \rho_-) = \eta$, whereas $D(\sigma_+, \sigma_-) = 1$; hence at least one of the two recovery errors is at least $(1 - \eta)/2$.
Universal single-copy recovery to arbitrary accuracy is therefore impossible in this setting, independently of the estimator.
What the estimators of this paper exploit is precisely the additional information that oracle access supplies---equivalent to many copies and a measurement scheme---and all resource claims must be read with that cost included.
We therefore refrain from claims of saturating or beating tomographic rates; Fig.~\ref{fig:sample_complexity} shows the empirical infidelity--copy curves of the two effective strategies at $d = 8$ against the $O(n^{-1})$ and $O(n^{-1/2})$ reference slopes, with the reference lines drawn as visual guides (their prefactors are illustrative, not derived), and the curves are flat: neither strategy consumes more than one state representation, so neither improves with $n$ within this framework.

\subsection{End-to-end application: VQE for H$_2$}
\label{sec:vqe_demo}

To demonstrate practical utility, we applied blind CQEC within a variational quantum eigensolver (VQE) for the H$_2$ ground state energy at bond length $0.735$\,\AA{} in the STO-3G basis ($d = 4$, two qubits).
At each VQE iteration, the ansatz state $|\psi(\theta)\rangle$ is subjected to combined noise ($\gamma = 0.5$, $p = 0.1$, $\gamma_\mathrm{AD} = 0.05$), and the energy is evaluated as $\langle H \rangle = \tr(H\rho)$ on the (possibly corrected) state.

\begin{figure}[!tb]
\centering
\includegraphics[width=\columnwidth]{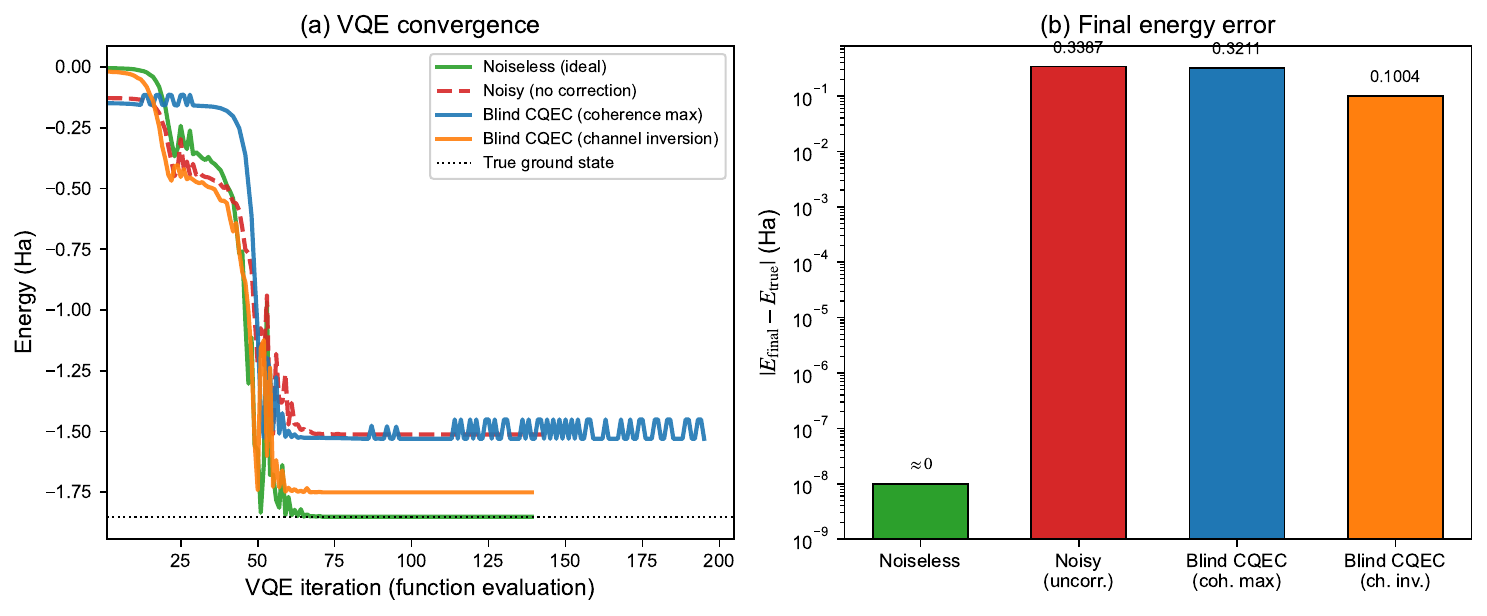}
\caption{\textbf{Channel-inversion blind CQEC reduces the noisy-VQE energy error for H$_2$ by $3.4\times$ without target-state knowledge.}
(a)~Energy vs.\ iteration for noiseless (green), noisy without correction (red), blind CQEC with coherence maximization (blue), and blind CQEC with channel inversion (purple); dashed line marks the exact ground state energy.
(b)~Final energy error $|E_\mathrm{opt} - E_\mathrm{exact}|$ for each scenario.}
\label{fig:vqe}
\end{figure}

Figure~\ref{fig:vqe} compares four scenarios: noiseless (ideal), noisy without correction, blind CQEC with coherence maximization, and blind CQEC with channel inversion.
The noiseless VQE converges to the exact ground state ($E_0 = -1.851$\,Ha).
Without correction, noise induces an energy bias of $\Delta E = 0.34$\,Ha.
Channel-inversion blind CQEC reduces this to $\Delta E = 0.10$\,Ha---a $3.4\times$ improvement---without any knowledge of the ansatz state at each iteration.
Coherence maximization provides a marginal improvement ($\Delta E = 0.32$\,Ha), reflecting the fact that VQE noise is dominated by amplitude damping (population relaxation), which coherence maximization cannot address (Sec.~\ref{sec:cohmax}).

This demonstrates that blind CQEC can provide tangible benefits in a quantum chemistry workflow, even when the target state evolves across optimization iterations and is never explicitly known.

For balance we also report a negative result from the same benchmark suite: an analogous VQE run for LiH at $d = 16$ failed to converge in \emph{all four} scenarios---including the noiseless one---with final energy errors of $\approx 6.2$\,Ha throughout, so blind CQEC neither helped nor harmed (the optimizer, not decoherence, was the bottleneck).
The H$_2$ result should therefore be read as a proof of principle on a case where the VQE itself converges, not as evidence that blind CQEC rescues arbitrary variational workflows; characterizing when correction helps in larger active spaces requires a systematically tuned optimizer study.

\subsection{Circuit-derived states: an estimator exercise with oracle calibration}
\label{sec:circuit_sanity}

All preceding results operate on states produced directly by our analytical noise channels.
As a complementary exercise, we applied the estimation strategies to density matrices produced by the \texttt{qiskit-aer} simulator (v0.17, with \texttt{qiskit} v2.4) for five small state-preparation circuits: GHZ states on $n = 2$ and $n = 3$ qubits, a $W$-like state, and two Haar-random parameterized circuits (\path{benchmark_blind_circuit_sanity.py}), each executed under a per-gate depolarizing noise model with $p_\mathrm{depol} = 0.10$ on single- and two-qubit gates.

Two scope limitations must be stated plainly, because they mean this exercise is \emph{not} a blind circuit-level validation.
First, the effective global depolarizing rate $p_\mathrm{eff}$ used by channel inversion is calibrated from the trace overlap with the \emph{ideal} circuit output---information a blind protocol would not possess; in a deployment, $p_\mathrm{eff}$ would come from independent calibration (e.g., randomized benchmarking~\cite{Emerson2005,Knill2008,Magesan2011}), and the values reported here are oracle-calibrated upper baselines.
Second, the reported fidelities are those of the \emph{estimates} against the ideal state; consistent with the recovery interface of Eq.~\eqref{eq:recovery_structure}, no catalytic circuit is executed.
What the exercise does show is that the estimators behave on circuit-derived noisy states (whose noise is a gate-local process, not our analytical channel) the same way they behave on channel-generated states.

\begin{table}[!tb]
\caption{Estimator exercise on \texttt{qiskit-aer} circuit-derived states: estimate fidelity against the ideal pure state for five small circuits under per-gate depolarizing noise ($p_\mathrm{depol} = 0.10$).
The channel-inversion parameter $p_\mathrm{eff}$ is \emph{oracle-calibrated} from the trace overlap with the ideal state (see text); the CI column is therefore an upper baseline, not a blind result.}
\label{tab:circuit_sanity}

\begin{tabular}{lcccc}
Test ($d$) & $F_\mathrm{noisy}$ & $F_\mathrm{CM}$ & $F_\mathrm{CI}$ & $p_\mathrm{eff}$ \\
\hline
GHZ (2q, $d{=}4$)        & 0.880 & 0.950 & \textbf{0.965} & 0.16 \\
GHZ (3q, $d{=}8$)        & 0.804 & 0.880 & \textbf{0.926} & 0.22 \\
$W$-like (2q, $d{=}4$)   & 0.822 & \textbf{0.984} & 0.970 & 0.24 \\
Random (2q, $d{=}4$)     & 0.746 & 0.885 & \textbf{0.972} & 0.34 \\
Random (3q, $d{=}8$)     & 0.615 & 0.830 & \textbf{0.877} & 0.44 \\
\end{tabular}

\end{table}

\begin{figure}[!tb]
\centering
\includegraphics[width=\columnwidth]{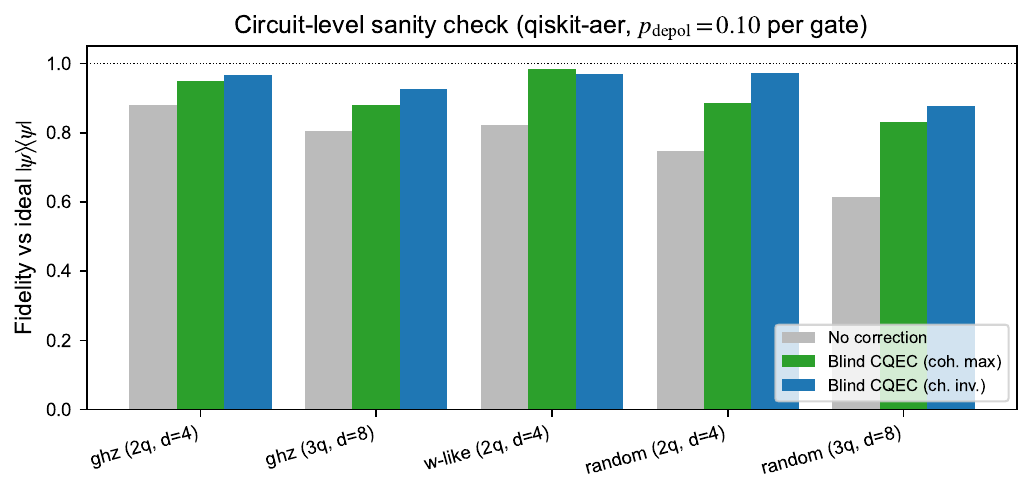}
\caption{\textbf{Estimator behavior on circuit-derived noisy states matches the channel-generated benchmarks; channel inversion here uses oracle-calibrated noise parameters.}
Five small state-preparation circuits under per-gate depolarizing noise ($p_\mathrm{depol} = 0.10$) on \texttt{qiskit-aer}.
Both estimators improve over the no-correction baseline in every case; the CI bars are oracle-calibrated upper baselines (see text).}
\label{fig:circuit_sanity}
\end{figure}

Table~\ref{tab:circuit_sanity} reports the resulting fidelities.
Both estimators improve the estimate over the no-correction baseline in every case: coherence maximization (fully blind) gains $5$--$22\%$, and oracle-calibrated channel inversion gains $9$--$26\%$.
The qualitative ordering of strategies is consistent with the density-matrix-level results at $d = 4, 8$ (Table~\ref{tab:depol_only}).
A genuine circuit-level validation of blind CQEC---with independently calibrated noise parameters and an executed catalytic recovery circuit---remains open and is listed in Sec.~\ref{sec:limitations}.

\section{Discussion}
\label{sec:discussion}

\subsection{Regimes of applicability}
\label{sec:regimes}

Our results delineate three operational regimes for blind CQEC:

\begin{enumerate}
\item \textbf{Low-dim, unknown noise, near-pure target} ($d \leq 16$, purity $v \gtrsim 0.7$): Coherence maximization is sufficient.
No noise-model information is needed, and a single state representation suffices ($\Frec \geq 0.986$ on the three algorithms with $d \leq 16$, Table~\ref{tab:main_results}).
The purity qualifier is essential: at $v \leq 0.7$ coherence maximization is \emph{worse than no correction} (Sec.~\ref{sec:mixed_state}), so a purity check must gate its use.
This regime covers most near-term quantum algorithms with nearly pure outputs (qDRIFT, QKAN, QPE).

\item \textbf{High-dim or mixed target, characterized noise}: Channel inversion is required.
Coherence maximization fails because population estimates become unreliable at high $d$ and because the physicality bound overshoots for mixed targets.
This regime requires noise characterization (e.g., via randomized benchmarking~\cite{Emerson2005,Knill2008,Magesan2011}) at the accuracy quantified in Sec.~\ref{sec:sensitivity}, but not target-state knowledge.

\item \textbf{Copy-constrained}: Channel inversion and coherence maximization consume one state representation and do not improve with additional copies; the strategies that do (iterative refinement, multi-copy averaging) need $n \gtrsim 100$ preparations to approach them (Table~\ref{tab:min_copies}), so under a copy budget the single-representation strategies are the only viable ones.
\end{enumerate}

A notable feature of the results is that \emph{naive estimation never improves over no correction}.
This occurs because CQEC is a directed recovery process that amplifies coherence modes selectively toward the target specification: when the target is set to the noisy state itself ($\rhoe = \rhon$), the noisy state is a fixed point of the recovery map, so no improvement occurs.
Multi-copy averaging of identical noisy copies likewise yields $\rhon$, so the naive strategy cannot be rescued by additional copies.

\subsection{Why coherence maximization degrades at high dimension}
\label{sec:cohmax_principle}

The success of coherence maximization at low $d$ and its degradation at high $d$ can be understood quantitatively.
For a $d$-dimensional maximally mixed noisy state, all populations approach $1/d$, and the physicality bound $\sqrt{p_i p_j} \approx 1/d$ carries vanishing information about the target's off-diagonal structure.
In contrast, at low $d$ with near-pure targets, the bound is tight and phase-preserving noise retains the exact phase structure (Sec.~\ref{sec:cohmax}), allowing near-perfect reconstruction.
The widening gap between coherence maximization and the recovery ceiling as $d$ grows thus reflects the transition from the information-rich to the information-poor regime of the population-based estimator; whether this translates into a preference for channel inversion in practice depends on the available calibration accuracy (Sec.~\ref{sec:sensitivity}).

\subsection{Comparison with standard QEC and purification-based suppression}
\label{sec:pqec}

It is important to position blind CQEC relative to standard stabilizer-based QEC~\cite{Shor1995,Gottesman1997}.
Standard QEC encodes logical qubits into physical qubits \emph{before} noise acts, enabling syndrome-based correction without knowledge of the encoded state.
This encoding-first paradigm is fundamentally different from CQEC, which operates \emph{after} noise has corrupted an unencoded state.
The two approaches are therefore not directly comparable: standard QEC requires foresight (pre-encoding), while blind CQEC is a post-hoc recovery method applicable when encoding was not performed or was insufficient.
Blind CQEC does not replace standard QEC but rather addresses a complementary scenario---recovering states from noisy quantum computations where pre-encoding was absent or where the noise exceeded the code's correction capacity.

Purification quantum error suppression (PQES)~\cite{Raghoonanan2026} also improves an unknown state without target knowledge, using SWAP-test purification of multiple noisy copies with the full outcome record retained (no postselection); that work reports error thresholds of $3/4$ (depolarizing) and $1/2$ (dephasing) for its scheme.
The key structural differences are:
\begin{itemize}
\item PQES uses recursive purification requiring multiple identical noisy copies, with the copy count growing with the desired precision;
blind CQEC, within the fluctuation-averaging model of Sec.~\ref{sec:copy_scaling}, needs no additional copies beyond those consumed by the estimator (single-representation estimators use one), with the caveat that oracle density-matrix access is assumed.
\item PQES operates without any noise-model information;
blind CQEC performs best when the noise model is known (channel inversion), though coherence maximization provides a noise-model-free alternative at low $d$.
\end{itemize}
The two approaches address complementary regimes: PQES is suited for settings where no noise information is available and copies are abundant; blind CQEC is suited for the copy-constrained, noise-characterized regime.

\subsection{Limitations}
\label{sec:limitations}

Several limitations warrant discussion; the first two are structural and bound the scope of every claim in this paper.
\begin{enumerate}
\item \emph{The catalytic transformation is not simulated.}
As stated in Secs.~\ref{sec:implementation} and~\ref{sec:correlation}, all benchmarks implement the density-matrix-level recovery interface [Eq.~\eqref{eq:recovery_structure}], which represents the idealized output of a perfectly executed catalytic covariant transformation.
The explicit CPTP map acting on $\rhon \otimes c$---with covariance, complete-positivity, and catalyst-marginal checks---is constructed in Ref.~\cite{Wakaura2026unified} for $d \leq 8$, where those checks show that the catalyst marginal is \emph{not} preserved (marginal fidelity $0.48$--$0.84$), and is not integrated into these benchmarks; consequently, all reported fidelities are upper bounds, and properties such as the estimation--recovery transparency of Sec.~\ref{sec:correlation} are properties of the interface, not verified properties of catalytic dynamics.
Independently of the interface, correlated catalysis carries a resource cost of its own: when the residual catalyst--system correlation must be small, the catalyst must be correspondingly resourceful~\cite{Rubboli2022limits}, a cost that the density-matrix-level accounting here does not include.
\item \emph{No operational measurement model.}
All estimators consume the exact noisy density matrix, which is not a physically accessible object at unit cost: acquiring it requires $\tilde\Theta(d^2/\epsilon^2)$ copies even with collective measurements (Sec.~\ref{sec:sample_complexity}).
Every resource count in this paper is conditional on this oracle access (Secs.~\ref{sec:qem_compare} and~\ref{sec:sample_complexity}), and the multi-copy ensembles use a Gaussian perturbation proxy rather than Born-rule statistics (Sec.~\ref{sec:copy_scaling}).
\item \emph{Noise-parameter sensitivity of channel inversion.}
The sensitivity analysis of Sec.~\ref{sec:sensitivity} shows a narrow, asymmetric tolerance window at high dimension: at $d = 64$, the advantage over the noise-model-free strategy survives only for calibration errors within roughly $[-5\%, +10\%]$, with the dephasing rate as the dominant (exponentially amplified) parameter; at $d \leq 8$, $\pm 10\%$ suffices.
\item \emph{Mixed-state targets.}
Coherence maximization becomes actively harmful below purity $v \approx 0.5$ (Sec.~\ref{sec:mixed_state}) and must be gated by a purity check; channel inversion is purity-robust.
\item \emph{Scalability and tensor product structure.}
The Haar-random sweep covers $d \leq 64$ as a single $d$-level system ($d \geq 128$ exceeds double-precision range for the exact dephasing inversion).
In realistic multi-qubit settings, noise acts locally, and estimation strategies could exploit this locality (e.g., per-qubit coherence maximization); extending the framework to tensor-product structure is a natural next step.
\item \emph{Experimental validation.}
All results are numerical; circuit-level implementation with independently calibrated noise and hardware validation remain open (the exercise of Sec.~\ref{sec:circuit_sanity} uses oracle-calibrated parameters and is not a substitute).
\item \emph{Dependence on companion preprints.}
The CQEC framework and its catalyst-preparation methods are given in our companion preprint~\cite{Wakaura2026unified} (arXiv:2603.25774, Secs.~2 and~4); the interpolation ansatz used by the legacy recovery code was first applied in the neural-spin-model study~\cite{Wakaura2026catalyst} (Research Square, doi:10.21203/rs.3.rs-9366084); the PQES scheme is Ref.~\cite{Raghoonanan2026} (arXiv:2603.11568).
Lemma~1, Theorem~2, Corollary~1, and the estimation strategies are self-contained, with every external fact cited in Sec.~\ref{sec:bottleneck}.
\end{enumerate}

\section{Computational Details}
\label{sec:computational}

All simulations were performed as density-matrix-level numerical calculations using custom Python code (NumPy~2.0.2, SciPy~1.13.1) on an Apple M4 Max workstation.
No quantum hardware was used; apart from the \texttt{qiskit-aer} exercise of Sec.~\ref{sec:circuit_sanity}, all operations act directly on $d \times d$ density matrices at cost $O(d^3)$ per recovery (Sec.~\ref{sec:implementation}); the full catalytic joint-system simulation, not performed here, would cost $O(d^6)$.

\emph{Unified noise module.}
All noise channels (forward and inverse) are implemented once, in \texttt{blind\char`\_cqec/noise.py} and \texttt{blind\char`\_cqec/estimators.py}, and every benchmark script imports these implementations; unit tests verify element-wise agreement between the scripts and the package (maximal deviation $\sim 10^{-16}$) and round-trip exactness of the inversion ($F > 0.999$ for $\EE^{-1}(\EE(\rho))$ at all $d \leq 64$).
The mode-inclusion threshold is fixed at $10^{-10}$ (matching the mode checker of the core library): coherences of $\rhon$ below this magnitude are treated as annihilated and are zeroed in the recovery target [Eq.~\eqref{eq:recovery_structure}]; this convention determines the $d = 64$ ceilings reported in Secs.~\ref{sec:algorithms_results} and~\ref{sec:haar_sweep}.
Table~\ref{tab:tau_sensitivity} quantifies the dependence: lowering $\tau_\mathrm{mode}$ lifts every ceiling, and at $\tau_\mathrm{mode} = 0$ (no mode restriction) exact channel inversion recovers each algorithm state perfectly, so the reported ceilings measure the pruning convention rather than a physical limit.
The PSD projection $\Pi$ also partially re-populates pruned entries: for the Regev state under combined noise with the oracle estimate, the $\ell_1$ weight that $\Pi$ places on positions zeroed by $\mathcal{M}_{\rhon}$ is $0.054$, i.e.\ $5.2\%$ of the output's total off-diagonal $\ell_1$ weight ($1.039$), which is why the two-stage map is described as conservative rather than as an exact mode filter.

\begin{table}[!tb]
\caption{Sensitivity of the exact-channel-inversion (oracle) recovery fidelity to the mode threshold $\tau_\mathrm{mode}$; $\tau_\mathrm{mode} = 0$ disables the mode restriction. Main-text values use $10^{-10}$.}
\label{tab:tau_sensitivity}
\begin{tabular}{lccccc}
Case & $10^{-10}$ & $10^{-12}$ & $10^{-14}$ & $10^{-16}$ & $0$ \\
\hline
Regev, combined            & 0.773 & 0.772 & 0.772 & 0.944 & 1.000 \\
Regev, dephasing only      & 0.411 & 0.411 & 0.411 & 0.772 & 1.000 \\
CF-QPE, dephasing only     & 0.709 & 0.755 & 1.000 & 1.000 & 1.000 \\
qDRIFT, combined           & 1.000 & 1.000 & 1.000 & 1.000 & 1.000 \\
\end{tabular}
\end{table}
Every figure and table is generated by one script and one results file, listed in \texttt{REPRODUCE.md} of the repository, with the raw numerical data shipped alongside.

All results are deterministic given the recorded seeds.
The target states are fixed by the choice of quantum algorithm (not randomly sampled) in the algorithm benchmarks, and Haar-random ensembles (20 states per dimension) are used for the dimension sweep, sensitivity, mixed-state, and hybrid benchmarks, with means and standard deviations reported.

The complete source code is publicly available at \url{https://github.com/deeptell-inc/blind_cqec_pkg}, including the reference implementation as a Python package (\texttt{blind\_cqec}, also published on PyPI) and all benchmark scripts:
\begin{itemize}\setlength{\itemsep}{0pt}\setlength{\parsep}{0pt}
  \item \path{benchmark_blind_cqec.py} -- qubit/qutrit noise sweeps;
  \item \path{benchmark_blind_dimension_sweep.py} -- Haar-random dimension sweep;
  \item \path{benchmark_blind_sensitivity.py} -- noise-parameter sensitivity;
  \item \path{benchmark_blind_tau_sensitivity.py} -- mode-threshold sensitivity (Table~\ref{tab:tau_sensitivity});
  \item \path{benchmark_blind_circuit_sanity.py} -- qiskit-aer circuit-level sanity check;
  \item \path{benchmark_blind_vqe_demo.py} -- end-to-end VQE demonstration for H$_2$;
  \item \path{benchmark_blind_algorithms.py}, \path{benchmark_blind_qem_compare.py}, and the figure-generation scripts.
\end{itemize}

\section{Conclusion}
\label{sec:conclusion}

We introduced blind catalytic quantum error correction: a two-stage protocol that removes the target-state knowledge requirement of CQEC by combining a classical pre-estimator with the density-matrix-level recovery interface of Ref.~\cite{Wakaura2026unified} (which that reference itself describes as target-parameterized, nonlinear, and not a CPTP channel).
We stated at the outset the structural fact that governs every result: at this level the recovery map reduces to a mode-inclusion-restricted projection of the estimate [Eq.~\eqref{eq:recovery_structure}], so recovery fidelity is bounded by estimation fidelity by construction ($\Frec \geq 1 - 2\|\rhoe - \rhot\|_1 - 2\Delta_\mathrm{mode}$ for density-matrix estimates, rigorously; Eq.~\eqref{eq:explicit_bound} in the general case).
The 84-condition benchmark confirms this bound is respected everywhere, reducing the design of blind CQEC---within this framework---to a classical estimation problem.

The benchmark's substantive content is the map of the estimator design space, which the unified noise implementation makes sharp.
With exactly calibrated parameters, channel inversion coincides with the oracle at every dimension, and the deficit that appears at $d \geq 32$ ($1.000 \to 0.297$ over $d = 2$--$64$ for Haar-random targets) is the recovery-interface ceiling set by the mode threshold, not estimation error; the noise-model-free alternative, coherence maximization, tracks the ceiling at low dimension and separates from it as populations approach $1/d$ ($0.979 \to 0.212$).
The operationally decisive variable is calibration accuracy: at $d = 64$, a $-10\%$ error in the noise parameters (dominated entirely by the dephasing rate) drops channel inversion below the noise-model-free strategy, and the sensitivity surface of Sec.~\ref{sec:sensitivity} quantifies exactly when each estimator should be preferred.
Two failure modes we report with equal prominence: coherence maximization is actively harmful for target purity $v \lesssim 0.7$, and no blind strategy reaches $\Frec = 0.95$ at $d = 64$ under combined noise.

These findings position blind CQEC as a candidate complement to the established error-handling stack---post-hoc and threshold-free where QEC is encoding-first and threshold-bounded, state-level where mitigation is observable-level---while the limitations analysis makes the gap to deployment explicit: the catalytic transformation itself is not simulated here, density-matrix access stands in for a measurement model whose tomographic cost dominates all resource counts, and the H$_2$ demonstration ($3.4\times$ energy-error reduction, Fig.~\ref{fig:vqe}) is a proof of principle accompanied by a null result on LiH.
The path from this proof of concept to a deployable protocol runs through three open problems, in order: an explicit CPTP catalytic recovery map evaluated on $\rhon \otimes c$ (extending the $d \leq 8$ construction of Ref.~\cite{Wakaura2026unified}, which does not yet preserve the catalyst, and integrating it end-to-end); an operational measurement model with explicit POVMs and shot budgets; and circuit-level validation with independently calibrated noise.
The estimator design-space map presented here---exact-inversion ceiling, noise-model-free fallback, calibration-accuracy crossover, and purity gating---is the part of the problem that is now settled at the interface level and will carry over to any such implementation.

\begin{acknowledgments}
Numerical simulations were performed using the scientific Python ecosystem (NumPy, SciPy, Matplotlib) on Apple Silicon hardware.
The reference implementation \texttt{blind\char`\_cqec} is released as an open-source Python package on PyPI under the MIT license.
\end{acknowledgments}

\section*{Data availability}
All simulation code, raw numerical data, and plotting scripts used to generate every figure and table in this paper are publicly available at \url{https://github.com/deeptell-inc/blind_cqec_pkg}. 
All results are deterministically reproducible with the fixed random seed \texttt{numpy.random.seed(42)}; the test suite shipped with the \texttt{blind\char`\_cqec} package locks in the headline numbers reported here.  
 
\bibliographystyle{apsrev4-2}
\bibliography{paper_blind_cqec_pra}  

\end{document}